\documentclass[oneside, 10pt, journal]{IEEEtran}
\usepackage{amsmath, amsthm, amssymb,algorithmic,algorithm,graphicx,subfig, mathtools, bbm}
\usepackage{tikz}
\usetikzlibrary{plotmarks}

\newtheorem{theorem}{\textbf{Theorem}}[section]
\newtheorem{assumption}{\textbf{Assumption}}
\newtheorem{definition}{\textbf{Definition}}[section]
\newtheorem{corollary}[theorem]{\textbf{Corollary}}
\newtheorem{lemma}[theorem]{\textbf{Lemma}}

\newtheoremstyle{example}{\topsep}{\topsep}%
 {}
 {}
 {\bfseries}
 {}
 {\newline}
 {\thmname{#1}\thmnumber{ #2}\thmnote{ #3}}

\theoremstyle{example}

\theoremstyle{plain}
\newtheorem*{theorem*}{\textbf{Theorem}}
\newtheorem*{lemma*}{\textbf{Lemma}}
\newtheorem*{corollary*}{\textbf{Corollary}}

\author{Nicholas Ruozzi and Sekhar Tatikonda
\thanks{N. Ruozzi is a member of the Communication Theory Laboratory at EPFL, and S. Tatikonda is a member of the Department of Electrical Engineering at Yale University.}
\thanks{This work was presented, in part, at the conference Uncertaintiy in Artificial Intelligence (UAI) 2010.}}

\date{}

\begin{document}
\title{Message-Passing Algorithms:  Reparameterizations and Splittings}
\maketitle

\begin{abstract}
The max-product algorithm, a local message-passing scheme that attempts to compute the most probable assignment (MAP) of a given probability distribution, has been successfully employed as a method of approximate inference for applications arising in coding theory, computer vision, and machine learning. However, the max-product algorithm is not guaranteed to converge to the MAP assignment, and if it does, is not guaranteed to recover the MAP assignment.  

Alternative convergent message-passing schemes have been proposed to overcome these difficulties.  This work provides a systematic study of such message-passing algorithms that extends the known results by exhibiting new sufficient conditions for convergence to local and/or global optima, providing a combinatorial characterization of these optima based on graph covers, and describing a new convergent and correct message-passing algorithm whose derivation unifies many of the known convergent message-passing algorithms.  

While convergent and correct message-passing algorithms represent a step forward in the analysis of max-product style message-passing algorithms, the conditions needed to guarantee convergence to a global optimum can be too restrictive in both theory and practice.  This limitation of convergent and correct message-passing schemes is characterized by graph covers and illustrated by example.

\end{abstract}
\begin{IEEEkeywords}
Graphical models, Maximum a posteriori estimation, Message passing, Belief propagation, Inference algorithms
\end{IEEEkeywords}

\section{Introduction}

Belief propagation was originally formulated by Judea Pearl to solve inference problems in Bayesian networks \cite{Pearl1982}.  Pearl demonstrated that, for tree-structured graphical models, a simple, distributed message-passing algorithm, dubbed ``belief propagation'', is guaranteed to converge to the exact marginals of the input
probability distribution.  If the belief propagation algorithm is run on an arbitrary graphical model (i.e., one that may contain cycles), then neither convergence nor correctness are guaranteed.  In practice, however, the ``loopy'' belief propagation algorithm often produces reasonable approximations to the true marginals \cite{empiricalbp}.

Pearl also proposed an algorithm for MAP estimation that he dubbed ``belief revision.''  This optimization analog of belief propagation is more commonly known as the max-product or, equivalently, the min-sum algorithm.  These algorithms have similar guarantees to the belief propagation algorithm:  they produce the correct MAP estimate when the graphical model is a tree and may or may not produce the correct solution when run on graphs that contain cycles.   In this work, we focus primarily on variants of the min-sum algorithm: a local message-passing scheme designed to find the global minimum of an objective function that can be written as a sum of functions, each of which depends on a subset of the problem variables.    

For arbitrary graphical models, the min-sum algorithm may fail to converge
\cite{malioutov}, it may converge to a set of beliefs from which a global minimum
cannot be easily constructed \cite{wainwright}, or the estimate extracted upon
convergence may not be optimal \cite{weisscomp}.  Despite these difficulties, the min-sum algorithm, like the belief propagation algorithm, has found empirical success
in a variety of application areas including
statistical physics, combinatorial optimization \cite{mwissanghavi}, computer
vision, clustering \cite{affinity}, and the minimization of quadratic functions
\cite{quadciamac} \cite{malioutov}.  However, rigorously characterizing the
behavior of the algorithm outside of a few well-structured instances has proved
challenging.  

In an attempt to improve the performance of the min-sum algorithm, recent work has produced alternative message-passing algorithms that, under certain conditions, are
provably convergent and correct:  MPLP \cite{MPLP}, serial tree-reweighted max-product (TRW-S)
\cite{kolserial}, max-sum diffusion (MSD) \cite{MSD}, and the norm-product \cite{hazan} algorithm.  These message-passing algorithms are convergent in the sense that they can each be viewed as coordinate-ascent schemes over concave lower bounds.  Such message-passing algorithms can be converted into distributed algorithms by performing multiple coordinate-ascent steps in parallel and then averaging the results.  Unfortunately, this process may require some amount of central control at each step and typically results in slower rates of convergence when compared with the original coordinate-ascent scheme \cite{sontag}.   A discussion of efficient parallel message passing based on the norm-product algorithm can be found in \cite{distconvex}.

The above algorithmic ideas are closely related to concurrent work in the coding community on pseudo-codewords and LP decoding \cite{becpseudo}, \cite{vontobel}.  Coordinate-ascent schemes and convergent message-passing algorithms related to linear programming problems that arise in the context of coding were studied in \cite{vont1}, \cite{vont2}, and \cite{vont3}.  As we will see, these approaches are connected to the above approaches via the MAP LP, a standard linear programming relaxation of the MAP problem.

Other related work has focused on convex free energy approximations and the convergence of the sum-product algorithm \cite{weissdbcnt}, \cite{heskes1}, \cite{heskes2}.  As the max-product algorithm is the zero temperature limit of the sum-product algorithm, these results provide an algorithmic alternative to the max-product algorithm but can suffer from numerical issues as a result of the limiting process.  Convex free energy approximations are constructed from a vector of double counting numbers.  These double counting numbers are closely related to the reparameterizations that we will consider in this work.

The primary focus of this work is to provide a systematic technique for the design of convergent and correct message-passing algorithms that, like the standard min-sum message-passing algorithm, are both decentralized and distributed.  Such algorithms have the potential to be useful for solving large-scale, parallel optimization problems for which standard optimization algorithms are impractical.

Our primary contributions are threefold:
\begin{itemize}
\item  We propose a new distributed, local message-passing algorithm, which we call the splitting algorithm, that contains many of the other convergent and correct message-passing algorithms as a special case.  This algorithm, though initially derived by ``splitting'' the nodes of the factor graph into multiple reweighted copies, can also be interpreted as producing an alternative reparameterization of the objective function.  We show how to derive an entire theory of message-passing algorithms by starting from reparameterizations (in contrast to much of the work on convergent message-passing algorithms that begins with Lagrangian duality).

\item We provide conditions under which this new algorithm can converge to locally and globally optimal solutions of the minimization problem.  Past work on similar message-passing algorithms has focused exclusively on global optimality.  Empirically, message-passing algorithms that do not guarantee the global optimality of extracted assignments may still perform better than their convergent and correct counterparts.  As such, understanding the behavior of message-passing algorithms that only guarantee certain forms of local optimality is of practical importance.

\item We characterize the precise relationship between graph covers and the convergence and correctness of local message-passing algorithms. This characterization applies not only to provably convergent and correct message-passing algorithms, but also to other message-passing schemes that only guarantee local optimality.  This understanding allows us to provide necessary and sufficient conditions for local message-passing algorithms to converge to the correct solution over pairwise binary graphical models.
\end{itemize}

Beyond these contributions, this work attempts to unify many disparate results in the theory of message-passing algorithms that have appeared across multiple communities as well as to make explicit the assumptions that are often used when studying graphical models.  This work is organized as follows.  In Section \ref{sec:prev}, we review the problem setting, the min-sum algorithm, and factor graphs.  In Sections \ref{sec:splitalg} and \ref{chp:reparam}, we derive the splitting algorithm from the min-sum algorithm and show how this result can be generalized to a study of reparameterizations.  In Section \ref{chp:opt}, we provide conditions under which the splitting algorithm is, in some sense, locally or globally optimal. In Section \ref{sec:convalgs}, we produce a simple, convergent message-passing algorithm from the splitting reparameterization via a trivial lower bound and coordinate ascent.  In Section \ref{sec:gcover}, we show how to understand the fixed-point solutions of the splitting algorithm in terms of graph covers and discuss the extension of the theory of pseudo-codewords to general message-passing algorithms.  In Section \ref{sec:lvg}, we provide examples that illustrate the limits of the convergent message-passing approach. Finally, in Section \ref{sec:conc}, we summarize our results and conclude.

\section{Previous Work}
\label{sec:prev}
In this section, we review the problem formulation, necessary terminology, and basic results concerning the min-sum algorithm.  Let $f: \prod_i\mathcal{X}_i \rightarrow \mathbb{R} \cup \{\infty\}$, where each
$\mathcal{X}_i$ is an arbitrary set (e.g., $\mathbb{R}$, $\{0,1\}$, $\mathbb{Z}$,
etc.). Throughout this paper, we will be interested in finding an element
$(x_1,\ldots, x_n)\in \prod_i\mathcal{X}_i$ that minimizes $f$, and as such, we
will assume that there is such an element:

\begin{assumption}
$\exists x^* \in\prod_i\mathcal{X}_i$ such that $f(x^*) = \inf_x f(x)$.
\end{assumption}

We note that we will in general allow $f$ to take the value $\infty$ over its domain.  However, as we will see in subsequent sections, some results will only apply when $f$ is a proper, real-valued function (i.e., $f$ does not take the value $\infty$ or $-\infty$ at any element in its domain).

For an arbitrary function, computing
this minimum may be computationally expensive, especially if $n$ is large.  A typical scientific application may involve hundreds of thousands of variables and potential functions, and storing the entire problem on one computer may be difficult, if not impossible.  In other applications, such as sensor networks, processing power and storage are limited.  Because local message-passing algorithms like the min-sum algorithm are decentralized and distributed, they can operate on scales at which typical algorithms would be impractical.  

Although we will discuss algorithms for the minimization problem, some applications have a more natural formulation as maximization problems.  In these instances, we can use the equivalence $\max_x f(x) = -\min_x \big[-f(x)\big]$ in order to convert the maximization problem into a minimization problem.  Historically, the max-product algorithm for nonnegative functions is often studied instead of the min-sum algorithm.  We prefer the min-sum algorithm for notational reasons, and all of the results discussed in this work can easily be converted into results for the max-product case.

\subsection{Factorizations and Factor Graphs}
\label{sec:factgraphs}
The basic observation of the min-sum algorithm is that, even though the original
minimization problem may be difficult, if $f$ can be written as a sum of
functions depending on only a small subset of the variables, then we may be able
to minimize the objective function by performing a series of minimizations over
(presumably easier) sub-problems.  To make this concrete, let $\mathcal{A}
\subseteq 2^{V}$.   We say that $f$ \textit{factorizes} over a hypergraph $G=(V,\mathcal{A})$ if
we can write $f$ as a sum of real valued potential functions $\phi_i :
\mathcal{X}_{i\in V} \rightarrow \mathbb{R} \cup \{\infty\}$ and $\psi_{\alpha} :
\mathcal{X}_{\alpha} \rightarrow \mathbb{R}\cup \{\infty\}$ as 
\begin{align}
f(x) = \sum_{i\in V} \phi_i(x_i) + \sum_{\alpha \in \mathcal{A}}
\psi_{\alpha}(x_\alpha). \label{f}
\end{align}

In this work, we focus on additive factorizations, and multiplicative factorizations, such that all of the potential functions are nonnegative, can be converted into additive factorizations by taking a negative log of the objective function.  

The above factorization is by no means unique.  For example, suppose we are given
the objective function $f(x_1,x_2) = x_1 + x_2 + x_1x_2$.  We can factorize $f$ in many
different ways.
\begin{align}
f(x_1,x_2) & =  x_1 + x_2 + x_1x_2\\
& =  x_1 + (x_2 + x_1x_2)\\
& =  (x_1 + x_2 + x_1x_2)\\
& =  x_1 + x_2 + \frac{x_1x_2}{2} + \frac{x_1x_2}{2} \label{hfac}
\end{align}
Each of these rewritings represents a different factorization of $f$ (the parenthesis indicate a single potential function).  All of these
factorizations can be captured by the above definitions, except for the last.  Recall
that $\mathcal{A}$ was taken to be a subset of $2^{V}$.  In order to
accommodate the factorization given by \eqref{hfac}, we will allow
$\mathcal{A}$ to be a multiset whose elements are members of the set $2^{V}$.

The set of all factorizations of the objective function $f(x)$ over $G = (V,\mathcal{A})$ forms an affine set, 
\begin{align}
\mathcal{F}_{(V,\mathcal{A})}(f) = &\{(\phi, \psi) : \kappa + \sum_{i\in V} \phi_i(x_i) + \sum_{\alpha \in \mathcal{A}} \psi_{\alpha}(x_\alpha) = f(x)\nonumber\\
&\: \text{for all $x$}\}.
\end{align}
If $(\phi, \psi) \in \mathcal{F}_{(V,\mathcal{A})}(f)$ and $(\phi', \psi') \in \mathcal{F}_{(V,\mathcal{A})}(f)$, then $(\phi, \psi)$ is called a \textit{reparameterization} of $(\phi', \psi')$ and vice versa.

\begin{figure}
\centering
\scalebox{.8}{
  \begin{tikzpicture}[scale=1.5]
	\tikzstyle{every node}=[draw,shape=circle];	
	\path (90:1) node (X0) {$x_1$};
	\path (210:1) node (X2) {$x_2$};	
	\path (330:1) node (X4) {$x_3$};
	\tikzstyle{every node}=[draw,shape=rectangle,minimum size=.75cm];		
	\path (150:1) node (X1) {$\psi_{12}$};
	\path (270:1) node (X3) {$\psi_{23}$};
	\path (30:1) node (X5) {$\psi_{13}$};
	
	\draw (X0) -- (X1);
	\draw (X1) -- (X2);
	\draw (X2) -- (X3);
	\draw (X3) -- (X4);
	\draw (X4) -- (X5);
	\draw (X5) -- (X0);
	\end{tikzpicture}
}
\caption[Example factor graph.]{The factor graph corresponding to $f(x_1,x_2,x_3) = \phi_1 + \phi_2 +
\phi_3 + \psi_{12} + \psi_{23} + \psi_{13}$.  By convention, variable nodes are
represented as circles and factor nodes are represented as squares.  Typically, the $\phi$ functions that depend only on a single variable are omitted from the graphical representation (for clarity).  }
\label{factfig}
\end{figure}
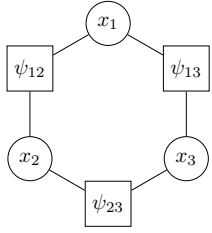

We can represent the hypergraph $G = (V,\mathcal{A})$ as a bipartite graph with a variable node $i$ for each variable $x_i$, a factor node $\alpha$ for each of the potentials $\psi_\alpha$, and an edge joining the factor node corresponding to $\alpha$ to the variable node
representing $x_i$ for all $i \in \alpha$.  This bipartite graph is called the factor graph representation of $G$.  Factor graphs provide a visual representation of the relationship among the potential functions.  In this work, we will always assume that $G$ is given in its factor graph representation.  For a concrete example, see Figure \ref{factfig}.    

\subsection{The Min-Sum Algorithm}
\label{sec:minsum}
The min-sum algorithm is a local message-passing algorithm over a factor graph.  During the execution of the min-sum algorithm, messages are passed back and forth between adjacent nodes of the graph.  In the algorithm, there are two types of messages: messages passed from variable nodes to factor nodes and messages passed from factor nodes to variable nodes.  On the $t^{th}$ iteration of the algorithm, messages are passed along each edge of the factor graph as follows:
\begin{align}
m^t_{i \rightarrow \alpha}(x_i) & \coloneqq   \kappa + \phi_i(x_i) + \sum_{\beta\in\partial i \setminus \alpha}  m^{t-1}_{\beta\rightarrow i}(x_i) \label{eq:msvarfact}\\
m^t_{\alpha\rightarrow i}(x_i) & \coloneqq  \kappa + \min_{x_{\alpha \setminus i}} \Big[\psi_\alpha(x_\alpha) + \sum_{k\in\alpha \setminus i} m^{t-1}_{k\rightarrow\alpha}(x_k)\Big] \label{eq:msfactvar}
\end{align}
where $\partial i$ denotes the set of all $\alpha \in \mathcal{A}$ such that $i\in \alpha$ (intuitively, this is the set of neighbors of variable node $x_i$ in the factor graph), $x_\alpha$ is the vector formed from the entries of $x$ by selecting only the indices in $\alpha$, and $\alpha \setminus i$ is abusive notation for the set-theoretic difference $\alpha \setminus \{i\}$.  When the graph is a tree, these message updates can be derived by dynamic programming.  When the graph is not a tree, the same updates are used as if the graph was a tree.  Understanding when these updates converge to the correct solution for a given graph is the central question underlying the study of the min-sum algorithm.

Each message update has an arbitrary normalization factor $\kappa$.  Because $\kappa$ is not a function of any of the variables, it only affects the value of the minimum and not where the minimum is located.  As such, we are free to choose it however we like for each message and each time step.  In practice, these constants are used to avoid numerical issues that may arise during the execution of the algorithm.  

\begin{definition}
A vector of messages, $m = (\{m_{\alpha\rightarrow i}\}, \{m_{i\rightarrow\alpha}\})$, is \textbf{real-valued} if for all $\alpha\in\mathcal{A}$, $\forall i\in\alpha$, and $\forall x_i \in\mathcal{X}$, $m_{\alpha \rightarrow i}(x_i)$ and $m_{i \rightarrow \alpha}(x_i)$ are real-valued functions (i.e., they do not take the value $\infty$ for any $x_i\in\mathcal{X}$).
\end{definition}

We will think of the messages as a vector of functions indexed by the direced edge over which the message is passed.  Any vector of real-valued messages is a valid choice for the vector of initial messages $m^0$, and the choice of initial messages can greatly affect the behavior of the algorithm.  A typical assumption is that the initial messages are chosen such that $m^0_{\alpha\rightarrow i} \equiv 0$ and  $m^0_{i\rightarrow \alpha} \equiv 0$.  This uniformity assumption is often useful when we need to analyze the evolution of the algorithm over time, but ideally, we would like to design message-passing schemes that perform well independent of initialization.

We can use the messages in order to construct an estimate of the min-marginals of $f$.  Recall that a min-marginal of $f$ is a function of one or more variables obtained by fixing a subset of the variables and minimizing the function $f$ over all of the remaining variables.  For example, the min-marginal for the variable $x_i$ would be the function $f_i(x_i) \triangleq \min_{x' : x'_i = x_i} f(x')$.  Given any vector of messages, $m^t$, we can construct a set of beliefs that are intended to approximate the min-marginals of $f$ by setting
\begin{align}
b^t_i(x_i) & \coloneqq  \kappa + \phi_i(x_i) + \sum_{\alpha \in \partial i} m^t_{\alpha \rightarrow i}(x_i)\label{varb}\\
b^t_{\alpha}(x_\alpha) & \coloneqq  \kappa + \psi_\alpha(x_\alpha) + \sum_{i \in \alpha} m^t_{i \rightarrow \alpha}(x_i).\label{factb}
\end{align}
where, again, $\kappa$ is an arbitrary constant that can be different for each belief.

If the beliefs corresponded to the true min-marginals of $f$ (i.e., $b^t_i(x_i) = \min_{x' : x'_i = x_i} f(x')$), then for any $y_i\in\arg\min_{x_i} b^t_i(x_i)$ there exists a vector $x^*$ such that $x^*_i = y_i$ and $x^*$ minimizes the function $f$.  If $|\arg\min_{x_i} b^t_i(x_i)| = 1$ for all $i$, then we can construct $x^*$ by setting $x^*_i = y_i$ for all $i$, but, if the objective function has more than one optimal solution, then we may not be able to construct such an $x^*$ so easily.  For this reason, theoretical results in this area typically assume that the objective function has a unique global minimum. Although this assumption is common, we will \emph{not} adopt this convention in this work. 

Because our beliefs are not necessarily the true min-marginals, we can only approximate the optimal assignment by computing an estimate of the argmin
\begin{align}
x_i^t & \in \arg \min_{x_i} b^t_i(x_i).
\end{align}

\begin{definition}
A vector, $b = (\{b_i\}, \{b_\alpha\})$, of beliefs is \textbf{locally decodable} to $x^*$ if for all $i$ and for all $x_i\neq x_i^*$, $b_i(x_i^*) < b_i(x_i).$  Equivalently, for all $i$, $b_i$ has a unique minimum at $x_i^*$.
\end{definition}

If the algorithm converges to a vector of beliefs that are locally decodable to $x^*$, then we hope that the vector $x^*$ is a global minimum of the objective function.  This is indeed the case when the factor graph contains no cycles.  Informally, this follows from the correctness of dynamic programming on a tree.  This result is well known, and we will defer a more detailed discussion of this result until later in this work (see Corollary \ref{cor:tree}).

Similarly, we hope that the beliefs constructed from any fixed point of \eqref{eq:msvarfact} and \eqref{eq:msfactvar} would be the true min-marginals of the function
$f$.  If the beliefs are the exact min-marginals, then the estimate
corresponding to our beliefs would indeed be a global minimum.  Unfortunately,
the algorithm is only known to produce the exact min-marginals on special factor
graphs (e.g., when the factor graph is a tree, see Section \ref{sec:admiss}).  Instead, we will show that the
fixed-point beliefs are similar to min-marginals.  Like the messages, we will
think of the beliefs as a vector of functions indexed by the nodes of the factor
graph.  Suppose $f$ factors over $G = (V, \mathcal{A})$.  Consider the following definitions.

\begin{definition}
\label{admis}
A vector of beliefs, $b$, is \textbf{admissible} for a function $f$ if 
\[f(x) = \kappa + \sum_i b_i(x_i) + \sum_\alpha \Big[b_\alpha(x_\alpha) -
\sum_{k\in\alpha} b_k(x_k)\Big]\]
for all $x$.  Beliefs satisfying this property are said to reparameterize the objective
function.
\end{definition}

\begin{definition}
\label{def:consis}
A vector of beliefs, $b$, is \textbf{min-consistent} if for all $\alpha$ and all
$i\in\alpha$,
\[\min_{x_{\alpha \setminus i}} b_\alpha(x_\alpha) = \kappa + b_i(x_i) \]
for all $x_i$.
\end{definition}

Any vector of beliefs that satisfies these two properties produces a
reparameterization of the original objective function in terms of the beliefs.  As the following theorem demonstrates, any
vector of beliefs obtained from a fixed point of the message updates in \eqref{eq:msvarfact} and \eqref{eq:msfactvar} satisfies these two properties.
\begin{theorem}
For any vector of fixed-point messages, the corresponding beliefs are admissible
and min-consistent.\label{fpts}
\end{theorem}

A proof of Theorem \ref{fpts} can be found in Appendix \ref{app:fpts}.  This result is not new, and simialr proofs can be found, for example, in \cite{wainwright}.  We present the proof in the appendix only for completeness, and we will make use of similar proof ideas in subsequent sections.

\section{The Splitting Algorithm}
\label{sec:splitalg}
In this section, we provide a simple derivation of a reweighted message-passing scheme that can be derived from the min-sum algorithm by ``splitting'' the factor and variable nodes in the factor graph.  This novel approach shows that reweighted message-passing schemes can be derived from the standard min-sum algorithm on a modified factor graph, suggesting a close link between these types of message-passing schemes and factorizations.  Although this construction appears to make the message-passing scheme more complicated, in subsequent sections, we will see that similar ideas can be used to derive convergent and correct message-passing schemes.

Suppose $f$ factorizes over $G = (V,\mathcal{A})$ as in \eqref{f}. Take one potential $\alpha\in\mathcal{A}$ and split it into $c$
potentials $\alpha_1,\ldots,\alpha_c$ such that for each $j\in\{1, \ldots, c\}$,
$\psi_{\alpha_j}(x_\alpha) = \frac{\psi_\alpha(x_\alpha)}{c}$ for all $x_\alpha$.  This allows us to rewrite the
objective function, $f$, as
\begin{align}
f(x) & =  \sum_{i\in V} \phi_i(x_i) + \sum_{\beta \in \mathcal{A}}
\psi_{\beta}(x_\beta)\\
& =  \sum_{i\in V} \phi_i(x_i) + \sum_{\beta \in \mathcal{A} \setminus \alpha}
\psi_{\beta}(x_\beta) + \sum_{j = 1}^c \frac{\psi_\alpha(x_\alpha)}{c}\\
& =   \sum_{i\in V} \phi_i(x_i) + \sum_{\beta \in \mathcal{A} \setminus \alpha}
\psi_{\beta}(x_\beta) + \sum_{j = 1}^c \psi_{\alpha_j}(x_\alpha).
\end{align}

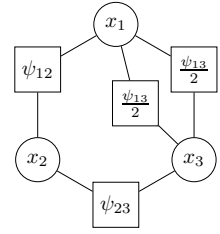
\begin{figure}
\centering
\scalebox{.8}{
  \begin{tikzpicture}[scale=1.5]
	\tikzstyle{every node}=[draw,shape=circle];	
	\path (90:1) node (X0) {$x_1$};
	\path (210:1) node (X2) {$x_2$};	
	\path (330:1) node (X4) {$x_3$};
	\tikzstyle{every node}=[draw,shape=rectangle,minimum size=.75cm];		
	\path (150:1) node (X1) {$\psi_{12}$};
	\path (270:1) node (X3) {$\psi_{23}$};
	\path (30:.25) node (X5) {$\frac{\psi_{13}}{2}$};
	\path (30:1) node (X6) {$\frac{\psi_{13}}{2}$};
	
	\draw (X0) -- (X1);
	\draw (X1) -- (X2);
	\draw (X2) -- (X3);
	\draw (X3) -- (X4);
	\draw (X4) -- (X5);
	\draw (X5) -- (X0);
	\draw (X4) -- (X6);
	\draw (X6) -- (X0);
	
	\end{tikzpicture}
	}
\caption[Splitting the factor nodes of a factor graph.]{The new factor graph formed by splitting the $\psi_{13}$ potential of the
factor graph in Figure \ref{factfig} into two potentials.}
\label{splitfactfig}
\end{figure}

This rewriting does not change the objective function, but it does produce a new
factor graph $F$ in which some of the variable nodes have a higher degree (see Figure \ref{splitfactfig}).  Now, take some $i\in \alpha$
and consider the messages $m_{i\rightarrow\alpha_j}$ and $m_{\alpha_j\rightarrow
i}$ given by the standard min-sum algorithm
\begin{align}
m^t_{i \rightarrow \alpha_j}(x_i) & =  \kappa + \phi_i(x_i) +
\sum_{\beta\in\partial_F i \setminus \alpha_j}  m^{t-1}_{\beta\rightarrow
i}(x_i)\\
m^t_{\alpha_j \rightarrow i}(x_i) & = \kappa+ \min_{x_{\alpha\setminus i}}\Big[
\frac{\psi_\alpha(x_\alpha)}{c} + \sum_{k\in \alpha_j \setminus i}  m^{t-1}_{k
\rightarrow \alpha_j}(x_k)\Big]
\end{align}
where $\partial_F i$ denotes the neighbors of $i$ in $F$.  Notice that there is
an automorphism of the graph that maps $\alpha_j$ to $\alpha_{j'}$ for all $j,j'\in\{1,\ldots,c\}$.  As the
messages passed from any node only depend on the messages received at the
previous time step, if the initial messages are the same at both of these nodes,
then they must produce identical messages at time 1. More formally, if we
initialize the messages identically over each split edge, then, at any time step
$t \geq 0$, $m^t_{i \rightarrow \alpha_j}(x_i) = m^t_{i \rightarrow
\alpha_{j'}}(x_i)$ for all $x_i$ and $m^t_{\alpha_j \rightarrow i}(x_i) = m^t_{\alpha_{j'} \rightarrow i}(x_i)$ for all $x_i$.  Because of this,
we can rewrite the message from $i$ to $\alpha_j$ as
\begin{align}
m^t_{i \rightarrow \alpha_j}(x_i)  =  &\: \phi_i(x_i) +
\sum_{\beta\in\partial_F i \setminus \alpha_j}  m^{t-1}_{\beta\rightarrow
i}(x_i)\\
= &\:  \phi_i(x_i) +  \sum_{l\neq j}m^{t-1}_{\alpha_l\rightarrow i}(x_i)\nonumber\\
&\:+ \sum_{\beta\in\partial_G i \setminus \alpha}  m^{t-1}_{\beta\rightarrow
i}(x_i)\\
= &\:  \phi_i(x_i) + (c-1)m^{t-1}_{\alpha_j\rightarrow i}(x_i)\nonumber\\
&\:{+}\sum_{\beta\in\partial_G i \setminus \alpha}  m^{t-1}_{\beta\rightarrow i}(x_i).
\label{rewrite}
\end{align}

Notice that \eqref{rewrite} can be viewed as a message-passing algorithm
on the original factor graph.  The primary difference then between the update in
\eqref{rewrite} and the min-sum update from \eqref{eq:msvarfact}, in addition to the scaling factor, is that the message passed from
$i$ to $\alpha$ now depends on the message from $\alpha$ to $i$.

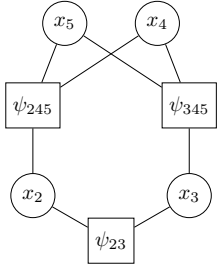
\begin{figure}
\centering
\scalebox{.8}{
  \begin{tikzpicture}[scale=1.5]
	\tikzstyle{every node}=[draw,shape=circle];	
	\path (70:1.5) node (X0) {$x_4$};
	\path (110:1.5) node (X6) {$x_5$};
	\path (210:1) node (X2) {$x_2$};	
	\path (330:1) node (X4) {$x_3$};
	\tikzstyle{every node}=[draw,shape=rectangle,minimum size=.75cm];		
	\path (150:1) node (X1) {$\psi_{245}$};
	\path (270:1) node (X3) {$\psi_{23}$};
	\path (30:1) node (X5) {$\psi_{345}$};
	
	\draw (X0) -- (X1);
	\draw (X1) -- (X2);
	\draw (X2) -- (X3);
	\draw (X3) -- (X4);
	\draw (X4) -- (X5);
	\draw (X5) -- (X0);

	\draw (X6) -- (X1);	
	\draw (X5) -- (X6);
	
	\end{tikzpicture}
	}
\caption[Splitting the variable nodes of a factor graph.]{New factor graph formed from the factor graph in Figure \ref{factfig}
by splitting the variable node $x_1$ into two variables $x_4$ and $x_5$.  The
new potentials are given by $\phi_4 = \phi_5 = \frac{\phi_1}{2}$, $\psi_{245} =
\psi_{12}(x_4, x_2) - \log \{x_4 = x_5\}$, and $\psi_{345} = \psi_{13}(x_4, x_3)
- \log \{x_4 = x_5\}$.}
\label{varsplitfig}
\end{figure}

Analogously, we can also split the variable nodes.  Suppose $f$ factorizes over
$G = (V,\mathcal{A})$ as in \eqref{f}.  Now, we will take one variable $x_i$ and split it into $c$
variables $x_{i_1},\ldots,x_{i_c}$ such that for each $l\in\{1,\ldots,c\}$, $\phi_{i_l}(x_{i_l}) = \frac{\phi_i(x_{i_l})}{k}$ for all $x_{i_l}$.  Again, this produces a new factor graph, $F$.  Because
$x_{i_1},\ldots,x_{i_k}$ are meant to represent the same variable, we must add a constraint to
ensure that they are indeed the same.  Next, we need to modify the potentials to
incorporate the constraint and the change of variables.  We will construct
$\mathcal{A}_F$ such that for each $\alpha\in\mathcal{A}$ with $i\in\alpha$
there is a $\beta = (\alpha\setminus i) \cup \{i_1,\ldots,i_c\}$ in
$\mathcal{A}_F$. Define $\psi_\beta(x_\beta) \triangleq \psi_\alpha(x_{\alpha\setminus
i}, x_{i_1}) - \log \mathbbm{1}_{x_{i_1} = \ldots = x_{i_c}}$ for all $x_{\alpha\setminus i}$.  For each
$\alpha\in\mathcal{A}$ with $i\notin\alpha$ we simply add $\alpha$ to
$\mathcal{A}_F$ with its old potential.  For an example of this construction,
see Figure \ref{varsplitfig}.  This rewriting produces a new objective function
\begin{align}
g(x) & = \sum_{j\neq i} \phi_j(x_j) + \sum_{l=1}^c \frac{\phi_i(x_{i_l})}{c} +
\sum_{\alpha \in \mathcal{A}_F} \psi_{\alpha}(x_\alpha). 
\end{align}
Minimizing $g$ is equivalent to minimizing $f$.  Again, we will show that we can
collapse the min-sum message-passing updates over $F$ to message-passing updates
over $G$ with modified potentials.  Take some $\alpha\in\mathcal{A}_F$
containing the new variable ${i_1}$ that augments the potential
$\gamma\in\mathcal{A}$ and consider the messages $m_{i_1\rightarrow\alpha}$ and
$m_{\alpha\rightarrow i_1}$ given by the standard min-sum algorithm
\begin{align}
m^t_{i_1 \rightarrow \alpha}(x_{i_1}) & =  \kappa + \frac{\phi_i(x_{i_1})}{c} +
\sum_{\beta\in\partial_F i_1 \setminus \alpha}  m^{t-1}_{\beta\rightarrow
i_1}(x_{i_1})\\
m^t_{\alpha \rightarrow i_1}(x_{i_1}) & =  \kappa + \min_{x_{\alpha\setminus
i_1}} \Big[\psi_\alpha(x_\alpha) + \sum_{k\in \alpha \setminus i_1}  m^{t-1}_{k
\rightarrow \alpha}(x_k)\Big].
\end{align}

Again, if we initialize the messages identically over each split edge, then, at
any time step $t \geq 0$, $m^t_{i_1 \rightarrow \alpha}(x_i) = m^t_{i_l
\rightarrow \alpha}(x_i)$ for all $x_i$ and $m^t_{\alpha \rightarrow i_1}(x_i) = m^t_{\alpha
\rightarrow i_l}(x_i)$ for all $x_i$ and for any $l \in\{1,\ldots,k\}$ by symmetry.  Using this, we
can rewrite the message from $\alpha$ to $i_1$ as
\begin{align}
m^t_{\alpha \rightarrow i_1}(x_{i_1}) = & \kappa + (c-1) m^{t-1}_{i_1 \rightarrow \alpha}(x_{i_1})\nonumber\\
& + \min_{x_{\alpha\setminus i_1}} \Big[\psi_\gamma(x_{\alpha\setminus i}, x_{i_1}) + \sum_{k\in \alpha \setminus i} m^{t-1}_{k \rightarrow \alpha}(x_k)\Big].
\end{align}

By symmetry, we only need to perform one message update to compute $m^t_{\alpha
\rightarrow i_l}(x_{i_l})$ for each $l\in\{1,\ldots,c\}$.  As a result, we can
think of these messages as being passed on the original factor graph $G$.  The
combined message updates for each of these splitting operations are described in Algorithm \ref{para}.  

\begin{algorithm*}[t]
\caption{Synchronous Splitting Algorithm\label{para}}
\begin{algorithmic}[1]
\STATE Initialize the messages to some finite vector.

\STATE For iteration $t = 1,2,\ldots$ update the the messages as follows
\begin{eqnarray*}
m^t_{i \rightarrow \alpha}(x_i) &  \coloneqq  & \kappa + \frac{\phi_i(x_i)}{c_i} +
(c_\alpha - 1)m^{t-1}_{\alpha\rightarrow i}(x_i) + \sum_{\beta\in\partial i
\setminus \alpha} c_\beta m^{t-1}_{\beta\rightarrow i}(x_i)\\
m^t_{\alpha\rightarrow i}(x_i) &  \coloneqq  & \kappa + \min_{x_{\alpha \setminus i}}
\Big[\frac{\psi_\alpha(x_\alpha)}{c_\alpha} + (c_i -
1)m^{t-1}_{i\rightarrow\alpha}(x_i) + \sum_{k\in\alpha \setminus i}
c_km^{t-1}_{k\rightarrow\alpha}(x_k)\Big].\label{eqmsgalpha}
\end{eqnarray*}
\end{algorithmic}
\end{algorithm*}

Throughout this discussion, we have assumed that each factor was split into $c$
pieces where $c$ was some positive integer.  If we allow $c$ to be an arbitrary
non-zero real, then the notion of splitting no longer makes sense.  Instead, as
described in Section \ref{sec:splits}, these splittings can be viewed more generally
as producing reparameterzations of the objective function.

\section{Reparameterizations}
\label{chp:reparam}

Recall from Theorem \ref{fpts} that the beliefs produced from fixed points of the min-sum algorithm are admissible: every vector of fixed-point beliefs, $b^*$, for the min-sum algorithm on the graph $G = (V,\mathcal{A})$ produces a reparameterization of the objective function
\begin{align}
f(x) & =  \sum_{i\in V} \phi_i(x_i) + \sum_{\alpha\in\mathcal{A}} \psi_\alpha(x_\alpha)\\
& =  \sum_{i\in V} b^*_i(x_i) + \sum_{\alpha\in\mathcal{A}} \Big[b^*_\alpha(x_\alpha) -\sum_{k\in\alpha} b^*_k(x_k)\Big].\label{eq:oldrp} 
\end{align}

In other words, we can view the min-sum algorithm as trying to produce a reparameterization of the objective function over $G$ in terms of min-consistent beliefs.  If the factor graph is a tree, then as was observed by Pearl and others, the min-marginals of the objective function produce such a factorization.  For each $i$ and $x_i$, let $f_i(x_i)$ be the min-marginal for the variable $x_i$, and for each $\alpha\in\mathcal{A}$ and $x_\alpha$, let $f_\alpha(x_\alpha)$ be the min-marginal for the vector of variables $x_\alpha$. 

\begin{lemma}
If $f$ factors over $G=(V,\mathcal{A})$ and the factor graph representation of $G$ is a tree, then $f$ can be reparameterized in terms of its min-marginals as
\begin{align}
f(x) & =  \kappa + \sum_{i\in V} f_i(x_i) + \sum_{\alpha\in\mathcal{A}} \Big[f_\alpha(x_\alpha) - \sum_{k\in\alpha} f_k(x_k)\Big].\label{eq:tree}
\end{align}
\end{lemma}
\begin{IEEEproof}
For example, see Theorem 1 of \cite{wainwright}.\\
\end{IEEEproof}

When the factor graph is not a tree, the min-marginals of the objective function do not necessarily produce a factorization of the objective function in this way, but we can still hope that we can construct a minimizing assignment from admissible and min-consistent beliefs.

In this section, we explore reparameterizations in an attempt to understand what makes one factorization of the objective function better than another.  Reparameterizations, and lower bounds derived from them, will be an essential ingredient in the design of convergent and correct message-passing algorithms.  In Section \ref{sec:admiss}, we use message reparameterizations to show that a slight modification to the definition of the beliefs for the min-sum algorithm can be used to ensure that beliefs corresponding to any vector of real-valued messages are admissible.  In Section \ref{sec:splits}, we show that a similar technique can be used to produce alternative reparameterizations of the objective function.  As in the case of the splitting algorithm, each of these reparameterizations will be characterized by a vector of non-zero reals.  These reparameterizations naturally produce lower bounds on the objective function.  Alternatively,  lower bounds can be derived using duality and a linear program known as the MAP LP \cite{sontag}. These two approaches produce similar lower bounds on the objective function.  In Section \ref{sec:maplp}, we review the MAP LP.

\subsection{Admissibility and the Min-Sum Algorithm}
\label{sec:admiss}

Fixed points of the message update equations produce a reparameterization of the objective function, but an arbitrary vector of messages need not produce a new factorization.  This difficulty is a direct consequence of having two types of messages (those passed from variables to factors and those passed from factors to variables).  However, we could ensure admissibility by introducing a vector of messages, $m'$, and rewriting the objective function as
\begin{align}
f(x) = &\: \sum_{i\in V} \Big[\phi_i(x_i) + \sum_{\alpha\in\partial i} m'_{\alpha\rightarrow i}(x_i)\Big]\nonumber\\
&\: + \sum_{\alpha\in\mathcal{A}} \Big[\psi_\alpha(x_\alpha) - \sum_{k\in\alpha} m'_{\alpha\rightarrow k}(x_k)\Big]. \label{streparam}
\end{align}
If the vector of messages is real-valued, this rewriting does not change the objective function.  For our new vector $m'$, consider the following definitions for the beliefs.
\begin{align}
b'_i(x_i) & =  \phi_i(x_i) + \sum_{\alpha \in \partial i} m'_{\alpha \rightarrow i}(x_i)\label{eq:nbel1}\\
b'_{\alpha}(x_\alpha) & =  \psi_\alpha(x_\alpha) + \sum_{k \in \alpha} \Big[b'_k(x_k) - m'_{\alpha\rightarrow k}(x_k)\label{eq:nbel2}\Big]
\end{align}
With these definitions, we can express the objective function as 
\begin{align}
f(x) = \sum_{i\in V} b'_i(x_i) + \sum_{\alpha\in\mathcal{A}} \Big[b'_\alpha(x_\alpha) - \sum_{k\in\alpha} b'_k(x_k)\Big].\label{eq:reparam}
\end{align}
Any choice of real-valued factor-to-variable messages produces an alternative factorization of the objective function.  Notice that the beliefs as defined in \eqref{eq:nbel1} and \eqref{eq:nbel2} only depend on one of the two types of messages.  Reparameterizations in the form of \eqref{eq:reparam} are meant to be reminiscent of the  reparameterization in terms of min-marginals for trees in \eqref{eq:tree}.  Notice that this definition of the beliefs corresponds exactly to those for the min-sum algorithm if we define
\begin{align} 
m'_{i\rightarrow \alpha}(x_i) & = b'_i(x_i) - m'_{\alpha\rightarrow k}(x_i)\\
& = \phi_i(x_i) + \sum_{\beta\in\partial i \setminus \alpha} m'_{\beta\rightarrow i}(x_i)
\end{align}
for all $\alpha$, all $i\in\alpha$, and all $x_i$.  

\subsection{The Splitting Reparameterization}
\label{sec:splits}
The min-sum algorithm produces reparameterizations of the objective function that have the same form as \eqref{eq:oldrp}.  Many other reparameterizations in terms of messages are possible.  For example, given a vector of non-zero reals, $c$, we can construct a reparameterization of the objective function,
\begin{align}
f(x) = &\: \sum_{i\in V} \Big[\phi_i(x_i) + \sum_{\alpha\in\partial i} c_ic_\alpha m_{\alpha\rightarrow i}(x_i)\Big]\nonumber\\
&\: + \sum_{\alpha\in\mathcal{A}} \Big[\psi_\alpha(x_\alpha) - \sum_{k\in\alpha} c_\alpha c_km_{\alpha\rightarrow k}(x_k)\Big]\\
 =&\:  \sum_{i\in V} c_i\Big[\frac{\phi_i(x_i)}{c_i} + \sum_{\alpha\in\partial i} c_\alpha m_{\alpha\rightarrow i}(x_i)\Big]\nonumber\\
&\: + \sum_{\alpha\in\mathcal{A}} c_\alpha\Big[\frac{\psi_\alpha(x_\alpha)}{c_\alpha} - \sum_{k\in\alpha} c_km_{\alpha\rightarrow k}(x_k)\Big]. \label{eq:newreparam}
\end{align}

By analogy to the min-sum algorithm, for each $i\in V$ and $\alpha\in\mathcal{A}$, we define the beliefs corresponding to this reparameterization as 
\begin{align}
b_i(x_i) & =   \frac{\phi_i(x_i)}{c_i} + \sum_{\alpha \in \partial i}
c_{\alpha}m_{\alpha \rightarrow i}(x_i)\label{varnew}\\
b_{\alpha}(x_\alpha) & =   \frac{\psi_\alpha(x_\alpha)}{c_\alpha} +
\sum_{k \in \alpha} c_k\Big[b_k(x_k) - m_{\alpha\rightarrow k}(x_k)\Big].\label{facnew}
\end{align}

This allows us to rewrite the objective function as
\begin{align}
f(x) = &\: \sum_{i\in V} c_ib_i(x_i)\nonumber\\
&\: + \sum_{\alpha\in\mathcal{A}} c_\alpha\Big[b_\alpha(x_\alpha) - \sum_{k\in\alpha} c_kb_k(x_k)\Big].
\end{align}

By analogy to the min-sum case, we will call any vector of beliefs that satisfies the above property for a particular choice of the parameters $c$-admissible.
\begin{definition}
\label{cadmis}
A vector of beliefs, $b$, is \textbf{$c$-admissible} for a function $f$ if 
\begin{align}
f(x) = &\: \kappa + \sum_{i\in V} c_ib_i(x_i)\nonumber\\
&\: + \sum_{\alpha\in\mathcal{A}} c_\alpha\Big[b_\alpha(x_\alpha) - \sum_{k\in\alpha} c_kb_k(x_k)\Big]
\end{align}
for all $x$. 
\end{definition}
Notice that if we choose $c_i = 1$ for all $i$ and $c_\alpha = 1$ for all $\alpha$ then we obtain the same reparameterization as the standard min-sum algorithm.  Because of the relationship to splitting the factors as in Section \ref{sec:splitalg}, we will call the reparameterization given by \eqref{eq:newreparam} the splitting reparameterization.  

We can use message reparameterizations to construct lower bounds on the objective function.  For example, consider the following lower bound obtained from the splitting reparameterization.
\begin{align}
\min_x f(x) \geq &\: \kappa + \sum_{i\in V} \min_{x_i} c_ib_i(x_i)\nonumber\\
&\: + \sum_{\alpha\in\mathcal{A}} \min_{x_\alpha}c_\alpha\Big[b_\alpha(x_\alpha) - \sum_{k\in\alpha} c_kb_k(x_k)\Big]
\end{align}

Notice that this lower bound is a concave function of the message vector, $m$, for any choice of the vector $c$ such that each component is nonzero.  Other concave lower bounds are possible using the same reparameterization
\begin{align}
\min_x f(x) \geq &\:  \kappa + \sum_{i\in V} \Big[\min_{x_i} c_i(1 - \sum_{\alpha\in\partial i}c_\alpha)b_i(x_i)\Big]\nonumber\\
&\: + \sum_{\alpha\in\mathcal{A}} \Big[\min_{x_\alpha}c_\alpha b_\alpha(x_\alpha)\Big].\label{eq:duallb}
\end{align}

\subsection{Lower Bounds and the MAP LP}
\label{sec:maplp}
\begin{figure*}[t]
\[
	\begin{array}{lrcl}
	\textrm{minimize}   & \sum_i \sum_{x_i} \mu_i(x_i)\phi_i(x_i) + \sum_\alpha \sum_{x_\alpha} \mu_\alpha(x_\alpha)\psi_\alpha(x_\alpha)&      &   \\
	\textrm{subject to} & \sum_{x_{\alpha\setminus i}} \mu_\alpha(x_\alpha) = \mu_i(x_i)    &     & \forall
\alpha\in\mathcal{A}, i\in\alpha, x_i\in\mathcal{X}_i\\
					& \sum_{x_i} \mu_i(x_i) = 1     &  &  \forall i\in V\\
			    & \mu_i(x_i) \in \{0,1\} & & \forall i, x_i\in\mathcal{X}_i\\
			    & \mu_\alpha(x_\alpha) \in \{0,1\} & & \forall \alpha\in\mathcal{A}, x_\alpha\in\prod_{i\in\alpha}\mathcal{X}_i
	\end{array}
\]
\caption{An integer programming formulation of the minimization problem corresponding to the factorization $(\phi,\psi)\in\mathcal{F}_\mathcal{A}(f)$. \label{maplp}}
\end{figure*}

Many authors have observed that, for finite state spaces (i.e., $|\mathcal{X}_i| < \infty$ for all $i$) and objective functions $f:\prod_i \mathcal{X}_i\rightarrow \mathbb{R}$, we can convert the optimization problem, $\min_x f(x)$, into an equivalent integer program by choosing a factorization$(\phi,\psi)\in\mathcal{F}_{(V,\mathcal{A})}(f)$ (see the definition in Section \ref{sec:factgraphs}) and introducing an indicator vector $\mu$ \cite{weissdbcnt, waintree}.  The resulting integer programming problem appears in Figure \ref{maplp}.  If $f$ is minimized at the assignment $x^*$, then choosing $\mu_i(x^*_i) = 1$ for all $i$, $\mu_\alpha(x_\alpha^*) = 1$ for all $\alpha$, and setting the remaining elements of $\mu$ to zero corresponds to an optimum of the integer program.

 If the objective function takes the value $\infty$ at some point $x\in\prod_i \mathcal{X}_i$, then, strictly speaking, the above construction is not technically an integer program.  We can correct for this by removing all infinite coefficients from the linear objective and forcing the corresponding $\mu$ variables to zero or one as appropriate.  As an example, if $\psi_\alpha(x_\alpha) = -\infty$ for some $x_\alpha$, then we will remove the $\mu_\alpha(x_\alpha)$ term from the linear objective and add a constraint that $\mu_\alpha(x_\alpha) = 0$ (this may result in the addition of exponentially many constraints).  

The integer program in Figure \ref{maplp} can be relaxed into a linear program by relaxing the integrality constraint, allowing $\mu_i(x_i)$ and $\mu_\alpha(x_\alpha)$ to be non-negative reals for all $i,x_i,\alpha,$ and $x_\alpha$.  The resulting linear program is typically referred to as the MAP LP.  We note that the constraints can be written in matrix form as $Ax =b$ such that the components of $A$ and $b$ are all integers.  Consequently, any vertex of the polytope corresponding to the system of equations $Ax = b$ must have all rational entries.

We can use the MAP LP and Lagrangian duality in order to construct lower bounds on the objective function; different duals will produce different lower bounds.  This approach produces lower bounds similar to the ones obtained in the last section and suggests a close relationship between duality and reparameterization.  Many different lower bounds on the objective function have been derived using duality (e.g., see \cite{MPLP}, \cite{sontag}, and \cite{waintree}).  For detailed discussions of duality and message-passing see \cite{sontag} and \cite{hazan}.  In addition, extensions of the duality arguments to continuous variable settings is discussed in \cite{continuousMAP}.

\section{Lower Bounds and Optimality}
\label{chp:opt}
In Section \ref{chp:reparam}, we saw that, given a vector of admissible beliefs, we can produce concave lower bounds on the objective function.  The discussion in this section will focus on the properties of admissible and min-consistent beliefs with relation to these lower bounds.  As such, these results will be applicable to a variety of algorithms, such as the min-sum algorithm, that produce beliefs with these two properties.  

Recall that, given a vector of admissible and min-consistent beliefs, we can, under certain conditions, construct a fixed-point estimate $x^*$ such that, for all $i$, $x_i^*\in\arg\min b_i$.  If the objective function had a unique global minimum and the fixed point
beliefs were the true min-marginals, then $x^*$ would indeed be the global
minimum.  Now, suppose that the $b_i$ are not the true min-marginals.  What can
we say about the optimality of a vector $x^*$ such that, for all $i$,  $x_i^* \in \arg\min
b_i$?  What can we say if there is a unique vector $x^*$ with this property?  We explore these questions by examining the lower bounds and reparameterizations discussed in Section \ref{chp:reparam}.  Our primary tool for answering these questions will be min-consistency and the following lemma.
\begin{lemma}
\label{uniq}
Let $b$ be a vector of min-consistent beliefs for a function $f$ that factorizes over $G=(V,\mathcal{A})$.  If $b$ is locally decodable to $x^*$, then 
\begin{itemize}
\item For all $\alpha\in\mathcal{A}$ and $x_\alpha$, $b_\alpha(x_\alpha) \geq b_\alpha(x^*_\alpha)$.
\item For all $\alpha\in\mathcal{A}$, $i\in \alpha$, and $x_\alpha$, $b_\alpha(x_\alpha) - b_i(x_i) \geq b_\alpha(x^*_\alpha) - b_i(x_i^*)$.
\end{itemize}
\end{lemma}
\begin{IEEEproof}
Because the beliefs are min-consistent for all $\alpha$ and any $i\in \alpha$, we have
\[\min_{x_{\alpha \setminus i}} b_\alpha(x_\alpha) = \kappa + b_i(x_i)\]
for all $x_i$ and some constant $\kappa$ that may depend on $i$ and $\alpha$ but does not depend on $x_i$.

From this, we can conclude that for each $\alpha\in\mathcal{A}$ and $i\in \alpha$ there is some $y_{\alpha}$ that minimizes
$b_\alpha$ with $y_i = x^*_i$.  Further, by the definition of locally decodable, the minimum is unique for each
$b_i$.  As a result, $x_\alpha^*$ must minimize $b_\alpha$.  Now fix a vector $x$ and consider
\begin{align}
b_\alpha(x^*_\alpha) - b_i(x^*_i) & =  \Big[\min_{x_{\alpha\setminus i}} b_\alpha(x_i^*,
x_{\alpha \setminus i})\Big] - b_i(x^*_i)\\
& =  \Big[\min_{x_{\alpha\setminus i}} b_\alpha(x_i, x_{\alpha \setminus i})\Big] - b_i(x_i)  \label{eq:constant}\\
& \leq  b_\alpha(x_{\alpha}) - b_i(x_i)
\end{align}
where \eqref{eq:constant} follows from the definition of min-consistency (this quantity is a constant independent of $x_i$).\\
\end{IEEEproof}

This lemma will be a crucial building block of many of the theorems in this
work, and many variants of this lemma have been proven in the literature (e.g.,
Lemma 4 in \cite{wainwright} and Theorem 1 in \cite{weissconv}). The lemma continues to hold, if the beliefs are not locally decodable, when there exists an $x^*$ that  simultaneously minimizes the beliefs in the following sense.

\begin{definition}
$x^*$ \textbf{simultaneously minimizes} a vector of beliefs, $b$, if $x^*_i\in\arg\min_{x_i} b_i(x_i)$ for all $i$ and $x^*_\alpha\in\arg\min_{x_\alpha} b_\alpha(x_\alpha)$ for all $\alpha$.
\end{definition}

If the beliefs are not locally decodable, we may not be able to efficiently construct an $x^*$ that simultaneously minimizes the beliefs (if one even exists).  Many of the results in subsequent sections will apply to any vector that simultaneously minimizes the beliefs, but we will focus on locally decodable beliefs for concreteness.

Using Lemma \ref{uniq} and admissibility, we can convert questions about the optimality
of the vector $x$ into questions about the choice of reparameterization.  Specifically, we focus on the splitting reparameterization
\begin{align}
f(x) = \sum_i c_ib_i(x_i) + \sum_\alpha c_\alpha\Big[b_\alpha(x_\alpha) - \sum_{k\in\alpha} c_kb_k(x_k)\Big].
\end{align}

In this section,  we will show how to choose the parameter vector $c$ in order to guarantee the local or global optimality of any estimate that simultaneously minimizes a vector of $c$-admissible and min-consistent beliefs.  

\subsection{Local Optimality}
\label{sec:locopt}
A function $f$ is said to have a local optimum at the point $x \in \prod_i\mathcal{X}_i$ if
there is some neighborhood of the point $x$ such that $f$ does not increase in that
neighborhood.  

The definition of neighborhood is metric dependent, and in the
interest of keeping our results applicable to a wide variety of spaces, we
choose the metric to be the Hamming distance.  For any two vectors $x,y
\in\prod_i\mathcal{X}_i$, the Hamming distance is the number of entries in which
the two vectors differ.  For the purposes of this paper, we will restrict our
definition of local optimality to vectors within Hamming distance one.

\begin{definition}
$x \in \prod_i\mathcal{X}_i$ is a \textbf{local minimum} of the objective
function, $f$, if for every vector $y$ that has at most one entry different from
$x$, $f(x) \leq f(y)$. \label{optdef}
\end{definition}

We will show that there exist choices of the parameters for which any estimate, extracted from a vector of $c$-admissible and min-consistent beliefs,
that simultaneously minimizes all of the beliefs is guaranteed to be locally
optimal with respect to the Hamming distance.  In order to prove such a result,
we first need to relate the minima of $c$-admissible and min-consistent beliefs to the minima of the objective function.  

Let $b$ be a vector of $c$-admissible beliefs for the function $f$.  Define $-j = \{1,\ldots,n\} \setminus \{j\}$.  For a fixed
$x_{-j}$, we can lower bound the optimum value of the objective function as
\begin{align}
\min_{x_j} f(x_j, x_{-j}) = &  \min_{x_j} \Bigg[\kappa + \sum_i c_ib_i(x_i)\nonumber\\
&\: +
\sum_\alpha c_\alpha\Big[b_\alpha(x_\alpha) - \sum_{k\in\alpha}
c_kb_k(x_k)\Big]\Bigg]\\
\geq &  \: g_j(x_{-j}) +  \Big[\min_{x_j} (1-\sum_{\alpha\in\partial
j}c_\alpha)c_jb_j(x_j)\Big]\nonumber\\
&\: + \sum_{\alpha \in \partial j} \Big[\min_{x_j} c_\alpha
b_\alpha(x_\alpha)\Big]\label{loceq}
\end{align}
where $g_j(x_{-j})$ is the part of the reparameterization that does not depend
on $x_j$.  The inequality is tight whenever there is a value of $x_j$ that
simultaneously minimizes each component of the sum.  If the coefficients of the
$b_j$'s and the coefficients of the $b_\alpha$'s in \eqref{loceq} were non-negative, then
we could rewrite this bound as 
\begin{align}
\min_{x_j} f(x) \geq & \: g_j(x_{-j}) +  \Big[(1-\sum_{\alpha\in\partial
j}c_\alpha)c_j  \min_{x_j} b_j(x_j)\Big]\nonumber\\
&\: + \sum_{\alpha \in \partial j} \Big[c_\alpha
\min_{x_j}b_\alpha(x_\alpha)\Big],
\end{align}
which depends on the minima of each of the beliefs.  Recall from Lemma
\ref{uniq} that if $b$ is locally decodable to $x^*$, then for all $i$ and $\alpha$, $x^*$ must simultaneously minimize  $b_i$,
$b_\alpha$, and, for $j\in \alpha$, $b_\alpha - b_j$.  So, in general, we want
to know if we can write 
\begin{align}
f(x) = & \: g_j(x_{-j}) + d_{jj}b_j(x_j) + \sum_{\alpha\in\partial j} d_{\alpha\alpha}
b_\alpha(x_\alpha)\nonumber\\
&\: + \sum_{\alpha\in\partial j} d_{j\alpha} \Big[b_\alpha(x_\alpha)
- b_j(x_j)\Big]
\end{align}
for each $j$ and some vector of non-negative constants $d$.  This motivates the following
definition.

\begin{definition}
\label{def:conical}
A function, $h$, can be written as a \textbf{conical combination} of the beliefs, $b$,
if there exists a vector of non-negative reals, $d$, such that
\begin{align} 
h(x) = & \: \kappa + \Big[\sum_{i,\alpha:i\in\alpha} d_{i\alpha}(b_\alpha(x_\alpha) -
b_i(x_i))\Big]\nonumber\\
&\: + \Big[\sum_\alpha d_{\alpha\alpha}b_\alpha(x_\alpha)\Big] + \Big[\sum_i d_{ii}
b_i(x_i)\Big].
\end{align}
\end{definition}
The set of all conical combinations of a collection of vectors in
$\mathbb{R}^n$ forms a cone in $\mathbb{R}^n$ in the same way that a convex
combination of vectors in $\mathbb{R}^n$ forms a convex set in $\mathbb{R}^n$.
  
The above definition is very similar to the definition of ``provably convex''
in \cite{weissconv}.  There, an entropy approximation is provably convex if it
can be written as a conical combination of the entropy functions corresponding
to each of the factors.  In contrast, our approach follows from a
reparameterization of the objective function. Putting all of the above ideas together, we have the following theorem.
\begin{theorem}
\label{loc}
Let $b$ be a vector of $c$-admissible and min-consistent beliefs for the function
$f$ such that for all $i$,  $c_ib_i(x_i) + \sum_{\alpha\in\partial i}
c_\alpha\Big[b_\alpha(x_\alpha) - c_ib_i(x_i)\Big]$ can be written as a conical
combination of the beliefs.  If the beliefs are locally decodable to $x^*$, then
$x^*$ is a local minimum (with respect to the Hamming distance) of the objective
function.  
\end{theorem}
\begin{IEEEproof}
Choose a $j \in \{1,\ldots,n\}$.  By assumption, the portion of the objective
function that depends on the variable $x_j$ can be written as a conical
combination of the beliefs.  By admissibility, up to a constant,
\begin{align}
 f(x^*) = &\: \sum_i c_ib_i(x^*_i) + \sum_\alpha
c_\alpha\Big[b_\alpha(x^*_\alpha) - \sum_{k\in\alpha} c_kb_k(x_k^*)\Big]\\
= &  \: g_j(x^*_{-j}) + c_jb_j(x_j^*)\nonumber\\
&\: + \sum_{\alpha \in \partial j}
c_\alpha\Big[b_\alpha(x^*_\alpha) - c_jb_j(x^*_j)\Big]\\
= & \: g_j(x^*_{-j}) + d_{jj}b_j(x^*_j) + \sum_{\alpha\in\partial j}
d_{\alpha\alpha} b_\alpha(x^*_\alpha) \nonumber\\
& \:+ \sum_{\alpha\in\partial j} d_{j\alpha}
[b_\alpha(x^*_\alpha) - b_j(x^*_j)]\\
\leq & \: f(x_j, x^*_{-j})
\end{align}
for any $x_j\in\mathcal{X}_j$.  The inequality follows from Lemma
\ref{uniq}, as $x^*$ simultaneously minimizes each piece of the sum, and the admissibility of $f$.  We can repeat this proof for each $j \in \{1,\ldots,n\}$.\\
\end{IEEEproof}

Theorem \ref{loc} tells us that, under suitable choices of the parameters, if the beliefs corresponding to a given fixed-point of the message-passing equations are locally decodable to $x^*$, then no vector $x$ within Hamming distance one of $x^*$ can decrease the objective function.  We can check that choosing $c_i = 1$ for all $i$ and $c_\alpha > 0$ for all $\alpha$ always satisfies the conditions of {Theorem~\ref{loc}}.  Consequently, if the fixed-point beliefs corresponding to the min-sum algorithm are locally decodable to $x^*$, then $x^*$ corresponds to a local optimum of the objective function.
\begin{corollary}
Let $b$ be a vector of admissible and min-consistent beliefs produced by the min-sum algorithm.  If the beliefs are locally decodable to $x^*$, then $x^*$ is a
local minimum (with respect to the Hamming distance) of the objective function.
\end{corollary}

Consider a differentiable function $f:\mathbb{R}^n\rightarrow \mathbb{R}$ (i.e., $\mathcal{X}_i = \mathbb{R}$ for all $i$).  Suppose that $c_i\neq 0$ for all $i$ and $c_\alpha \neq 0$ for all $\alpha$.  If a vector of $c$-admissible and min-consistent beliefs is locally decodable to $x^*$, then we can always infer that the gradient of
$f$ at the point $x^*$ must be zero (this is a direct consequence of min-consistency).  Consequently, $x^*$ is either a local minimum, a local maximum, or a saddle point of $f$.  If, in addition, $c$ satisfies the conditions Theorem \ref{loc}, then by the second derivative test and the observation that the function can only increase in value along the
coordinate axes, $x^*$ is either a local minimum or a saddle point of $f$.  Similarly, for a convex differentiable function $f:\mathbb{R}^n\rightarrow \mathbb{R}$, if $c_i\neq 0$ for all $i$ and $c_\alpha \neq 0$ for all $\alpha$, then $x^*$ minimizes $f$.

\begin{corollary}
\label{convexloc}
Let  $f:\mathbb{R}^n\rightarrow \mathbb{R}$ be a convex differentiable function. Under the hypothesis of Theorem \ref{loc}, if the beliefs are locally decodable to $x^*$, then $x^*$ is a global minimum of the objective function.  
\end{corollary}

\subsection{Global Optimality}
\label{sec:globopt}
We now extend the approach of the previous section to show that there are
choices of the vector $c$ that guarantee the global optimality of any unique
estimate produced from $c$-admissible and min-consistent beliefs.  As before,
suppose $b$ is a vector of $c$-admissible beliefs for the function $f$.  If $f$ can
be written as a conical combination of the beliefs for some vector $d$, then we can lower bound the
optimal value of the objective function as
\begin{align}
\min_x f(x) \geq & \sum_{i,\alpha:i\in\alpha} d_{i\alpha} \min_{x_\alpha}(b_\alpha(x_\alpha) - b_i(x_i))\nonumber\\
&\: + \sum_\alpha d_{\alpha\alpha}\min_{x_\alpha}b_\alpha(x_\alpha)\nonumber\\
&\: + \sum_i d_{ii} \min_{x_i}b_i(x_i).
\end{align}
This analysis provides us with our first global optimality result.  We note that
the following theorem also appears as Theorem 1 in \cite{weissconv}, and Theorem
2 in \cite{waintree} provides a similar proof for the TRMP algorithm.
\begin{theorem}
\label{correct2}
Let $b$ be a vector of $c$-admissible and min-consistent beliefs for the function
$f$ such that $f$ can be written as a conical combination of the beliefs.  If the beliefs
are locally decodable to $x^*$, then $x^*$ minimizes the objective function.
\end{theorem}

The proof of Theorem \ref{correct2} follows from Lemma  \ref{uniq} in nearly the same way as Theorem \ref{loc}.  The difference between
Theorem \ref{loc} and Theorem \ref{correct2} is that the former only requires
that the part of the reparameterization depending on a single variable can be
written as a conical combination of the beliefs, whereas the latter requires the
entire reparameterization to be a conical combination of the beliefs.  We can easily check that the conditions of Theorem \ref{correct2} imply the conditions of Theorem \ref{loc}, and as a result, Theorem \ref{correct2} places greater restrictions on the vector $c$.

As a corollary, Theorem \ref{correct2} also
provides us with a simple proof of the well-known result that the standard min-sum algorithm is correct
on a tree.

\begin{corollary}
\label{cor:tree}
Suppose the factor graph is a tree.  If the admissible and min-consistent
beliefs produced by the standard min-sum algorithm are locally decodable to
$x^*$, then $x^*$ is the global minimum of the objective function.
\end{corollary}
\begin{IEEEproof}
Let $b$ be the vector of min-consistent and admissible beliefs obtained from
running the standard min-sum algorithm.  Choose a variable node $r \in G$ and consider
the factor graph as a tree rooted at the variable node, $r$.  Let $p(\alpha)$
denote the parent of factor node $\alpha\in G$.  Because $G$ is a tree, we can now write, by admissibility, 
\begin{align}
f(x) & =  \kappa + \sum_{i} b_i(x_i) + \sum_{\alpha\in\mathcal{A}}
\Big[b_{\alpha}(x_\alpha) - \sum_{k\in\alpha}b_k(x_k)\Big]\\
& =  \kappa +  b_r(x_r) + \sum_{\alpha\in\mathcal{A}} \Big[b_{\alpha}(x_\alpha) -
b_{p(\alpha)}(x_{p(\alpha)})\Big].
\end{align}

Therefore, $f$ can be written as a conical combination of the
beliefs, and we can apply Theorem \ref{correct2} to yield the desired result.
\end{IEEEproof}

We note that there are choices of the parameters for which we are
guaranteed local optimality but not global optimality.    The standard min-sum algorithm always guarantees local optimality, and there are applications for which the algorithm is known to produce local optima that are not globally optimal \cite{weisscomp}.

Given Theorem \ref{correct2}, starting with the vector $d$ seems slightly more
natural than the starting with the vector $c$.  Consider any non-negative real
vector $d$, we now show that we can find a vector $c$ such that $f$ has a conical
decomposition in terms of $d$ provided $d$ satisfies a mild condition.  Given $d$, we will choose the vector $c$ as
\begin{align}
c_\alpha & = d_{\alpha\alpha} + \sum_{i\in \alpha} d_{i\alpha}\\
c_i & = \frac{d_{ii} - \sum_{\alpha\in\partial i} d_{i\alpha}}{1 -
\sum_{\alpha\in\partial i} c_\alpha}.
\end{align}

These equations are valid whenever ${1 - \sum_{\alpha\in\partial i}
c_\alpha}\neq 0$.  Note that any valid reparameterization must have $c_i \neq 0$
and $c_\alpha \neq 0$ for all $\alpha$ and $i$.  Hence, $d_{\alpha\alpha} +
\sum_{i\in \alpha} d_{i\alpha} \neq 0$ and  $d_{ii} \neq \sum_{\alpha\in\partial
i} d_{i\alpha}$.  

In the case that ${1 - \sum_{\alpha\in\partial i} c_\alpha} = 0$, $c_i$ can be
chosen to be any non-zero real.  Again, any valid reparameterization must have
$c_i \neq 0$ and $c_\alpha \neq 0$ for all $\alpha$ and $i$.  Hence,
$d_{\alpha\alpha} + \sum_{i\in \alpha} d_{i\alpha} \neq 0$, but, unlike the
previous case, we must have $d_{ii} - \sum_{\alpha\in\partial i} d_{i\alpha} =
0$.  

We now address the following question:  given a factorization of the objective
function $f$, does there exist a choice of the vector $c$ which
guarantees that any estimate obtained from locally decodable beliefs minimizes the objective function?  The answer to this question is yes,
and we will provide a simple condition on the vector $c$ that will ensure this. 

\begin{corollary}
\label{cor:correct}
Let $b$ be a vector of $c$-admissible and min-consistent beliefs for the function
$f$ such that
\begin{enumerate}
\item For all $i$, $(1 - \sum_{\alpha\in\partial i} c_\alpha)c_i \geq 0$
\item For all $\alpha$, $c_\alpha > 0$.
\end{enumerate}
If the beliefs are locally decodable to $x^*$, then $x^*$ minimizes the
objective function.
\end{corollary}
\begin{IEEEproof}
By admissibility, we can write $f$ as
\begin{align}
f(x) = &\: \sum_i \Big[(1-\sum_{\alpha\in\partial
i}c_\alpha)c_ib_i(x_i)\Big]\nonumber\\
&\: + \sum_\alpha \Big[c_\alpha
b_\alpha(x_\alpha)\Big].
\end{align}
Observe that if $(1-\sum_{\alpha\in\partial i}c_\alpha)c_i \geq 0$ for all $i$
and $c_\alpha \geq 0$ for all $\alpha$, then the above rewriting provides the desired conical decomposition of $f$.\\
\end{IEEEproof}

This result is quite general; for any choice of $c$ such that $c_\alpha > 0$ for
all $\alpha \in \mathcal{A}$, there exists a choice of $c_i$, possibly negative, for each $i\in V$ such
that the conditions of the above theorem are satisfied.  

All of the previous theorems are equally
valid for any vector that simultaneously minimizes all of the beliefs.  Given the above results, we would like to understand when the conditions of the theorems can be achieved.  Specifically, when can we guarantee that the algorithm converges to $c$-admissible and min-consistent beliefs that are locally decodable?  The remainder of this work attempts to provide an answer to this question.

\section{Convergent Algorithms}
\label{sec:convalgs}
\begin{algorithm*}[t]
\caption{Sequential Splitting Algorithm\label{serial}}
\begin{algorithmic}[1]
\STATE Initialize the messages uniformly to zero.

\STATE Choose some ordering of the variables, and perform the following update
for each variable $j$

\FOR{each edge $(j, \alpha)$}
\STATE For all $i\in\alpha \setminus j$, update the message from $i$ to $\alpha$
\[m_{i \rightarrow \alpha}(x_i) \coloneqq  \kappa + \frac{\phi_i(x_i)}{c_i} + (c_\alpha -
1)m_{\alpha\rightarrow i}(x_i) + \sum_{\beta\in\partial i \setminus \alpha}
c_\beta m_{\beta\rightarrow i}(x_i). \]
\STATE Update the message from $\alpha$ to $j$
\[m_{\alpha\rightarrow j}(x_j) \coloneqq \kappa + \min_{x_{\alpha \setminus j}}
\Big[\frac{\psi_\alpha(x_\alpha)}{c_\alpha} +  (c_j - 1)m_{j\rightarrow\alpha}(x_j)
+ \sum_{k\in\alpha \setminus j} c_k m_{k\rightarrow\alpha}(x_k)\Big].\]
\ENDFOR
\end{algorithmic}
\end{algorithm*}

Given that the lower bounds discussed in Sections \ref{chp:reparam} and \ref{chp:opt} are concave functions of the messages, we can employ traditional methods from convex optimization in an attempt to maximize them.  One such method is cyclic coordinate ascent. This scheme operates by fixing an initial guess for the solution to the optimization problem and then constructs a better solution by performing an optimization over a single variable.  However, this scheme does not always converge to an optimal solution.  For example, if the function concave is but not strictly concave, the algorithm may become stuck at local optima (again, a local optimum with respect to the Hamming distance, see Section \ref{sec:locopt}).  Despite this and other drawbacks, we will attempt to maximize our lower bounds on the objective function via block coordinate ascent, a variant of coordinate ascent where the update is performed over larger subsets of the variables at a time.  

Our proof of convergence will demonstrate that the proposed algorithm cannot decrease the lower bound (i.e., the value of the lower bound converges) and that, once the lower bound cannot be increased by further iterations, the beliefs behave as if they are min-consistent.  This definition of convergence does not guarantee that the beliefs or the messages converge, only that the lower bound converges.  First, we will discuss a particular coordinate-ascent scheme, and then we will discuss conditions under which the lower bound is guaranteed to be maximized by this and related schemes.

\subsection{A Simple Convergent Algorithm}
Consider the message-passing schedule in Algorithm \ref{serial}.  This
asynchronous message-passing schedule fixes an ordering on the variables and for
each $j$, in order, updates all of the messages from each $\beta \in\partial j$
to $j$ as if $j$ were the root of the subtree containing only $\beta$ and its
neighbors.  We will show that, for certain choices of the parameter vector $c$,
this message passing schedule cannot decrease a specific lower bound of the objective
function at each iteration.

To demonstrate convergence of the algorithm, we restrict the parameter vector $c$ so that $c_i = 1$ for all $i$, $c_\alpha >0$ for all $\alpha$, and $\sum_{\alpha \in\partial i} c_\alpha \leq 1$ for all $i$.  For a fixed vector of messages, $m$, consider the lower bound on the objective function
\begin{align}
\min_x f(x) \geq &\: \sum_i  \Big[c_i(1 - \sum_{\alpha\in\partial i}c_\alpha)\min_{x_i}b_i(x_i)\Big]\nonumber\\
&\: + \sum_\alpha \Big[c_\alpha \min_{x_\alpha}b_\alpha(x_\alpha)\Big],\label{eq:easylb}
\end{align}
where $b$ is the vector of beliefs derived from the vector of messages $m$.

Define $LB(m)$ to be the lower bound in \eqref{eq:easylb} as a function of the vector of messages, $m$.  We will show that, with this restricted choice of the parameter vector, Algorithm \ref{serial} can be viewed as a block coordinate
ascent scheme on the lower bound $LB(m)$.  In order to do so, we need the
following lemma.

\begin{lemma}
\label{serialcon}
Suppose $c_i = 1$ for all $i$, and we perform the update for the edge $(j, \alpha)$ as in Algorithm
\ref{serial}.  If the vector of messages is real-valued\footnote{This is a technical assumption.  Should there exist an edge $(i,j)$ such that for some $x_j$, $m_{ij}(x_j) \in \{\infty, -\infty\}$, then admissibility no longer makes sense.  This cannot occur for bounded functions on finite domains, but can occur, for example, when using this algorithm to minimize a multivariate quadratic function \cite{allerton09}.} after the update, then
$b_\alpha$ is min-consistent with respect to $b_j$.
\end{lemma}
\begin{IEEEproof}
See Appendix \ref{app:serialcon}.\\
\end{IEEEproof}

Observe that, after updating all of the factor-to-variable messages to a fixed variable node $j$ as in
Algorithm \ref{serial}, $b_\alpha$ is min-consistent with respect to $b_j$ for
every $\alpha\in\mathcal{A}$ containing $j$.  The most important conclusion we
can draw from this is that there is an $x_j^*$ that simultaneously minimizes
$b_j$, $\min_{x_{\alpha\setminus j}} b_\alpha$, and $\min_{x_{\alpha\setminus j}}
b_\alpha - b_j$.  

\begin{theorem}
\label{conv}
Suppose $c_i = 1$ for all $i$, $c_\alpha > 0$ for all $\alpha$, and
$\sum_{\alpha\in\partial i}c_\alpha \leq 1$ for all $i$.  If the vector of
messages is real-valued after each iteration of Algorithm \ref{serial}, then for all $t > 0$, $LB(m^t) \geq LB(m^{t-1})$.  \end{theorem}

The proof of Theorem \ref{conv} can be found in Appendix \ref{app:conv}.  We say that the lower bound has converged if for all $t > t'$, $LB(m^{t'}) = LB(m^t)$.  Again, this says nothing about the convergence of the messages or the beliefs.  However, as part of the proof of Theorem \ref{conv}, we show that for all $t > t'$, if $b^{t'}$ is locally decodable to $x^*$, then $x^*$ minimizes the objective function.

Although the restriction on the parameter vector in Theorem~\ref{conv} seems
strong, we observe that for any objective function $f$, we can choose the
parameters such that the theorem is sufficient to guarantee convergence and
global optimality.  As an example, if we set $c_\alpha = \frac{1}{\max_i |\partial i|}$ for all $\alpha$ and $c_i = 1$ for all $i$, then $c$ satisfies the conditions of Theorem \ref{conv}.

\begin{algorithm*}[t]
\caption{Damped Synchronous Splitting Algorithm\label{synchron}}
\begin{algorithmic}[1]
\STATE Fix a real-valued vector of initial messages, $m^0$.
\STATE Choose $\delta\in[0,1]$.

\FOR{$t = 1,2,\ldots$}
\STATE For each $\alpha$ and $i\in\alpha$, update the message from $i$ to $\alpha$,
\[m^t_{i \rightarrow \alpha}(x_i) \coloneqq  \kappa + \frac{\phi_i(x_i)}{c_i} + (c_\alpha -
1)m^{t-1}_{\alpha\rightarrow i}(x_i) + \sum_{\beta\in\partial i \setminus \alpha}
c_\beta m^{t-1}_{\beta\rightarrow i}(x_i). \]

\STATE For each $\alpha$ and $i\in\alpha$, update the message from $\alpha$ to $i$,
\[m^t_{\alpha\rightarrow i}(x_i) \coloneqq \kappa + (1-\delta) m^{t-1}_{\alpha\rightarrow i}(x_i)  + \delta\min_{x_{\alpha \setminus i}}
\Big[\frac{\psi_\alpha(x_\alpha)}{c_\alpha} +  (c_i - 1)m^t_{i\rightarrow\alpha}(x_i)
+ \sum_{k\in\alpha \setminus i} c_k m^t_{k\rightarrow\alpha}(x_k)\Big].\]
\ENDFOR
\end{algorithmic}
\end{algorithm*}
We present Algorithm \ref{serial} both as a convergent local message-passing algorithm and
as an example of how the intuition developed from the optimality conditions can
be extended to show convergence results: we can achieve global consistency by
carefully ensuring a weak form of local consistency.  Recall that, like other coordinate-ascent schemes, this algorithm can become stuck (i.e., reach a fixed point) that does not maximize the lower bound.  For an example of an objective function and choice of parameters such that the algorithm may become stuck, see \cite{kolserial}.

\subsection{Synchronous Convergence}
\label{snycconv}
By using the message-passing schedule in Algorithm \ref{serial}, we seem to lose
the distributed nature of the parallel message updates.  For some
message-passing schedules, we can actually parallelize the updating process by
performing concurrent updates as long as the simultaneous updates do not form a
cycle (e.g., we could randomly select a subset of the message updates that do not
interfere).  We also note that performing updates over larger subgraphs may be
advantageous, and other algorithms, such as those discussed in \cite{waintree}, \cite{kolserial}, \cite{weissconv}, and \cite{sontag}, perform updates over larger subtrees of the factor graph.

Each of these coordinate ascent schemes can be converted into distributed algorithms by performing multiple coordinate-ascent updates in parallel and then averaging the resulting message vectors.  Unfortunately, this process may require some amount of central control at each step and typically results in slower rates of convergence when compared with the original asynchronous message-passing scheme \cite{sontag}. For example, consider a message vector $m$.  Let $m^i$ be the vector of messages produced by performing the update for the variable node $i$ on the message vector $m$ as in Algorithm \ref{serial}.  For an appropriate choice of the vector $c$, Theorem \ref{conv} guarantees that $LB(m^i) \geq LB(m)$ for all $i$.  Since the lower bound is concave, we must also have that $LB(\sum_i \frac{m^i}{n}) \geq LB(m)$ where $n$ is the total number of variable nodes.  Let $m' = \sum_i \frac{m^i}{n}$. For all $\alpha$, $i\in\alpha$, and $x_i$,
\begin{align}
m'_{\alpha\rightarrow i}(x_i) = \frac{n-1}{n} m_{\alpha\rightarrow i}(x_i) + \frac{1}{n} m^i_{\alpha\rightarrow i}(x_i).
\end{align}

This observation can be used to construct the synchronous algorithm described in Algorithm \ref{synchron}.  The messages passed by this scheme are a convex combination of the previous time step and the splitting updates, modulated by a ``damping'' coefficient $\delta$.  Similar damped message updates are often employed in order to help the min-sum algorithm to converge.  Algorithm \ref{synchron} is guaranteed to converge when the parameter vector satisfies the conditions of Theorem \ref{conv} and $\delta = \frac{1}{n}$.  Other choices of $\delta$ can also result in convergence.  

\subsection{Relationship to Other Algorithms}
Recent work has produced other message-passing algorithms that are
provably convergent under specific updating schedules:  MPLP \cite{MPLP}, serial tree-reweighted max-product (TRW-S)
\cite{kolserial}, max-sum diffusion (MSD) \cite{MSD}, and the norm-product algorithm \cite{hazan}.  Like Algorithm \ref{serial}, these asynchronous message-passing algorithms are convergent in the sense that they can each be viewed as coordinate-ascent schemes over concave lower bounds.  

All of these algorithms, with the exception of the norm-product algorithm, were shown to be members of a particular family of bound minimizing algorithms \cite{weissconv}.  We note that, even when the parameter vector satisfies the conditions of Theorem \ref{conv},
Algorithm \ref{serial} is still not strictly a member of the family of bound
minimizing algorithms.  The disparity occurs
because the definition of a bound minimizing algorithm as presented therein
would require $b_\alpha$ to be min-consistent with respect to $x_i$ for all
$i\in\alpha$ after the update is performed over the edge $(j,\alpha)$.  Instead,
Algorithm \ref{serial} only guarantees that $b_\alpha$ is min-consistent with
respect to $x_j$ after the update.

In this section, we show that all of these message-passing algorithms can be seen as coordinate-ascent schemes over concave lower bounds.  More specifically, their derivations, with, perhaps, the exception of the norm-product algorithm, can be seen to follow the same formula developed earlier in this work: 
\begin{enumerate}
\item Choose a reparameterization.
\item Construct a lower bound.
\item Perform coordinate ascent in an attempt to maximize the bound.
\end{enumerate}
In some cases, the message-passing algorithms themselves can be seen as a special case of the splitting algorithm (although the original derivation of these algorithms was typically quite different than that of the splitting algorithm).  While in other cases, a slight tweak to the definition of min-consistency allows us to apply the results of the previous sections.

\subsubsection{TRW-S and TRMP}

The tree-reweighted belief propagation algorithm (TRBP) was first proposed in \cite{waintrbp}, and the application of similar ideas to the MAP inference problem is known as the tree-reweighted max-product algorithm (TRMP) \cite{waintree}.  At the heart of the min-sum analog of the TRMP algorithm is the observation that the objective function can be bounded from below by a convex combination of functions that depend only on factor induced subtrees of the factor graph.  As we will see below, the message updates of the TRMP algorithm, as defined in \cite{waintree}, are a special case of the splitting algorithm.  

Although the TRMP algorithm can be derived for general factor graphs, for simplicity, we consider the algorithm on a pairwise factor graph.  When the factorization is pairwise, each factor node is connected to at most two variable nodes.  As a result, the hypergraph $G= (V,\mathcal{A})$ can be viewed as a typical graph.  Each edge of $G$ corresponds to a factor node in the factor graph, and we will write $G=(V,E)$ in this case.

Let $\mathcal{T}$ be the set of all spanning trees on $G$, and let $\mu$ be a probability distribution over
$\mathcal{T}$ such that every edge has a nonzero probability of occurring in at
least one spanning tree.  Set $c_i = 1$ for all $i$ and $c_{ij} =
Pr_\mu[(i,j) \in T]$ corresponding to the edge appearance probabilities.  Let
$b$ be a vector of $c$-admissible and min-consistent beliefs for $f$.  We can write
the objective function $f$ as 
\begin{align}
f(x) = &  \sum_{i\in V} b_i(x_i)\nonumber\\
&\: + \sum_{(i,j)\in E} c_{ij}\Big[b_{ij}(x_i,x_j) - b_i(x_i) - b_j(x_j)\Big]\\
= & \sum_{T\in\mathcal{T}} \mu(T)\Bigg[\sum_{i\in V_T} b_i(x_i)\nonumber\\
&\: + \sum_{(i,j)\in
E_T} \Big[b_{ij}(x_i, x_j) - b_i(x_i) - b_j(x_j)\Big]\Bigg]
\end{align}
where $T =(V_T, E_T)$ is a spanning tree of $G$.

For each
$T\in\mathcal{T}$, designate a variable node $r_T \in T$ as the root of $T$. 
For all $T\in\mathcal{T}$ and $i\in T$, let $p^T(i)$ denote the parent of node $i\in T$.  We can now
write,
\begin{align}
f(x) = &\: \sum_{T\in\mathcal{T}} \mu(T)\Bigg[b_{r_T}(x_{r_T})\nonumber\\
&\: + \hspace{-.3cm}\sum_{i\in V_T, i\neq
r_T} \Big[b_{ip^T(i)}(x_i, x_{p^T(i)}) - b_{p^T(i)}(x_{p^T(i)})\Big]\Bigg]
\label{decomp}
\end{align}

Because $\mu(T) \geq 0$ for all $T\in\mathcal{T}$, we can conclude that $f$ can
be written as a conical combination of the beliefs.  The TRMP update, defined in \cite{waintree}, is then exactly Algorithm \ref{para} with the vector $c$ chosen as above.  All of the results from the previous sections can then be applied to this special case.  For example, by Theorem \ref{correct2},
convergence of the TRMP algorithm to locally decodable beliefs implies correctness.  

The TRMP algorithm was motivated, in part, by the observation that the min-sum algorithm is correct on trees.  However, a similar derivation can be made if $\mu$ is a probability distribution over all spanning subgraphs of
$G$ containing at most one cycle.  In this case, we would obtain a reparameterization of the objective function as a convex combination of functions over subgraphs containing only a single cycle.

Although the TRMP algorithm guarantees correctness upon convergence to locally decodable beliefs, the algorithm need not converge, even if we use Algorithm \ref{serial}.  Specifically, the vector $c$ does not necessarily satisfy the conditions of Theorem \ref{conv}.  The solution, proposed in \cite{kolserial}, is to perform the message updates over subtrees in a specific order that is guaranteed to improve the lower bound.  The resulting algorithm, known as the TRW-S algorithm, is then a convergent version of the TRMP algorithm.

\subsubsection{MPLP}
The MPLP algorithm was originally derived by constructing a special dual of the MAP LP from which a concave lower bound can be extracted.  The MPLP algorithm is then a coordinate-ascent scheme for this concave lower bound.  The MPLP algorithm was initially derived in terms of pairwise factor graphs, and was extended, with some work, to arbitrary factor graphs in \cite{MPLP}.  Unlike the TRMP algorithm, the MPLP algorithm does not have the tunable parameters required by the TRMP algorithm.  

Again, consider a pairwise factor graph, $G = (V,E)$, with corresponding objective
function, $f$.  Let $c_i = \frac{1}{2}$ for all $i$ and $c_{ij} = 1$ for all $i$.  This choice of $c$ produces the following reparameterization.
\begin{align}
f(x) = &\: \sum_{i\in V} \frac{b_i(x_i)}{2}\nonumber\\
&\: + \frac{1}{2}\sum_{(i,j)\in E}
\Big[b_{ij}(x_i,x_j) - b_i(x_i)\Big] \nonumber\\
&\:+ \frac{1}{2}\sum_{(i,j)\in E} \Big[b_{ij}(x_i,x_j) - b_j(x_j)\Big]\label{MPLPcon}
\end{align}
From \eqref{MPLPcon}, we can see that this choice of $c$ produces a
conical decomposition.  

Several variants of the MPLP algorithm were presented in the original paper.  One such variant, the EMPLP algorithm, can be seen as a coordinate-ascent scheme on the following lower bound.
\begin{align}
\min_x f(x) \geq &\: \sum_{i\in V} \min_{x_i}\frac{b_i(x_i)}{2}\nonumber\\
&\: + \frac{1}{2}\sum_{(i,j)\in E}
\min_{x_i, x_j}\Big[b_{ij}(x_i,x_j) - b_i(x_i)\Big]\nonumber\\
&\:+ \frac{1}{2}\sum_{(i,j)\in E}
\min_{x_i,
x_j}\Big[b_{ij}(x_i,x_j) - b_j(x_j)\Big]
\end{align}

We can rewrite the message update in terms of messages passed directly between the nodes of $G$.  Consider unwrapping the message updates as
\begin{align}
m^t_{(i,j)\rightarrow j}(x_j) = & \min_{x_j} \Big[\psi_{ij}(x_i, x_j)\nonumber\\
&\: -\frac{1}{2} m^{t-1}_{j\rightarrow (i,j)}(x_j)\nonumber\\
&\: +\frac{1}{2} m^{t-1}_{i\rightarrow (i,j)}(x_i)\Big]\\
= & \min_{x_i} \Bigg[\psi_{ij}(x_i, x_j)\nonumber\\
&\: -\frac{1}{2}\Big[\phi_j(x_j) + \sum_{k\in\partial j \setminus i} m^{t-1}_{(j,k)\rightarrow j}(x_j)\nonumber\Big]\\
&\: +\frac{1}{2}\Big[\phi_i(x_i) + \sum_{k\in\partial i \setminus j} m^{t-1}_{(i,k)\rightarrow i}(x_i)\Big]\Bigg].
\end{align}
Now, define 
\begin{align}
\widehat{m}^t_{i\rightarrow j}(x_j) \triangleq &\: \frac{1}{2}m^{t}_{(i,j)\rightarrow j}(x_j)\\
= & \min_{x_i} \Bigg[\frac{1}{2}\psi_{ij}(x_i, x_j)\nonumber\\
&\: -\frac{1}{2}\Big[\frac{1}{2}\phi_j(x_j) + \sum_{k\in\partial j \setminus i} \widehat{m}^{t-1}_{k\rightarrow j}(x_j)\nonumber\Big]\\
&\: +\frac{1}{2}\Big[\frac{1}{2}\phi_i(x_i) + \sum_{k\in\partial i \setminus j} \widehat{m}^{t-1}_{k\rightarrow i}(x_i)\Big]\Bigg]
\end{align}
for each $(i,j)\in E$.  This is precisely the message update for the EMPLP algorithm in \cite{MPLP} (there the self-potentials are identically equal to zero).  As was the case for the TRMP algorithm, we can extend all of the previous results to the general, non-pairwise, case. 

\subsubsection{Max-Sum Diffusion}
The max-sum diffusion algorithm and a variant known as the augmenting DAG algorithm were designed to solve the max-sum problem (i.e., the negated version of the min-sum problem).  Although discovered in the 1970s by Ukrainian scientists, most of the original work on these algorithms remained either in Russian or unpublished until a recent survey article \cite{MSD}. The augmenting DAG algorithm was suggested in \cite{augdag1} and later expanded in \cite{augdag2}.  The max-sum diffusion algorithm was discovered independently by two authors \cite{msd1, msd2}, but neither result was ever published.

Here, we derive the min-sum analog of the max-sum diffusion algorithm using the machinery that we have developed for the splitting algorithm.  Although the algorithm is a coordinate-ascent scheme over a familiar lower bound, the message updates are not an instance of the splitting algorithm because the fixed points are not min-consistent in the sense of Definition \ref{def:consis}.

The max-sum diffusion algorithm was originally described only for pairwise factorizations.  However, we will see that the algorithm can be derived for general factor graphs.  Consider the reparameterization of the objective function corresponding to the standard min-sum algorithm (i.e., $c_i = 1$
for all $i$ and $c_\alpha = 1$ for all $\alpha$)
\begin{align}
f(x) = &\:  \sum_i b_i(x_i) + \sum_{\alpha} \Big[b_\alpha(x_\alpha)
-\sum_{k\in\alpha} b_k(x_k)\Big]\\
= &\: \sum_i \min_{x_i} b_i(x_i)\nonumber\\
&\: + \sum_{\alpha} \min_{x_\alpha}
\Big[\psi_\alpha(x_\alpha) -\sum_{k\in\alpha} m_{\alpha\rightarrow k}(x_k)\Big].
\end{align}
The following lower bound follows from this reparameterization.
\begin{align}
\min_x f(x) \geq &\:  \sum_i \min_{x_i} b_i(x_i)\nonumber\\
&\: + \sum_{\alpha} \min_{x_\alpha}
\Big[\psi_\alpha(x_\alpha) -\sum_{k\in\alpha} m_{\alpha\rightarrow k}(x_k)\Big] \label{eq:msdlb}
\end{align}

The max-sum diffusion algorithm is a coordinate ascent message-passing scheme that improves the
above lower bound.  Unlike the reparameterizations that produced the TRMP and MPLP algorithms,
whether or not this reparameterization can be written as a conical combination
of the beliefs depends on the underlying factor graph.  As such, even if we choose an algorithm that converges to a min-consistent vector of beliefs, we will not be guaranteed correctness.  

Instead, the max-sum diffusion algorithm ensures a different form of consistency.  Namely, the algorithm guarantees that the fixed points of the message-passing scheme satisfy the following for each $\alpha$ and each $i\in\alpha$:
\begin{align}
\min_{x_{\alpha\setminus i}} \Big[b_\alpha(x_\alpha) - \sum_{k\in\alpha} b_k(x_k)\Big] = b_i(x_i). \label{eq:newconsis}
\end{align}

Again, there are many message updates that will guarantee this form of consistency upon convergence.  The one chosen by the developers of the max-sum diffusion algorithm was
\begin{align}
m_{\alpha\rightarrow i}(x_i) \coloneqq\: & m_{\alpha\rightarrow i}(x_i)\nonumber\\
&\: + \frac{1}{2}\min_{x_{\alpha\setminus i}} \Big[b_\alpha(x_\alpha) - \sum_{k\in\alpha} b_k(x_k) - b_i(x_i)\Big].
\end{align}
We can obtain a simpler message update rule that does not depend on the previous iteration as
\begin{align}
m_{\alpha\rightarrow i}(x_i) \coloneqq &\:  \frac{1}{2}\min_{x_{\alpha\setminus i}} \Big[\psi_\alpha(x_\alpha) - \sum_{k\in\alpha\setminus i} m_{\alpha\rightarrow k}(x_k)\Big]\nonumber\\
&\: - \frac{1}{2}\Big[\phi_i(x_i) - \sum_{\beta\in\partial i\setminus \alpha} m_{\beta\rightarrow i}(x_i)\Big].
\end{align}

After computing $m_{\alpha\rightarrow i}(x_i)$ for each $x_i$, the lower bound can only increase.  Further, we can check that if the algorithm converges to locally decodable beliefs, then this estimate is guaranteed to be correct.  This follows by replacing our notion of min-consistency with that of \eqref{eq:newconsis}.  In addition, the lower bound \eqref{eq:msdlb}, can be shown to be dual to the MAP LP \cite{MSD}. 

\subsubsection{Norm-Product}
The norm-product algorithm, like the above algorithms, is a coordinate-ascent scheme for maximizing a concave dual objective function \cite{hazan}.  Unlike the previous algorithms, however, whether or not the norm-product algorithm produces a reparameterization of the objective function remains an open question.

The algorithm is derived by studying the general problem of minimizing a convex objective function having a particular form.  The derivation of the algorithm uses more or less standard tools from convex analysis including Fenchel and Lagrangian duality. While the derivation of this algorithm is beyond the scope of this work, it is worth noting that, like the splitting algorithm, the norm-product algorithm is parameterized by a real vector.  For some choices of the parameter vector for both algorithms, the norm-product algorithm agrees with the asynchronous splitting algorithm (Algorithm \ref{serial}). 

\subsubsection{Subgradient Methods}

The fixed points of the splitting algorithm do not necessarily correspond to maxima of the lower bound.  As discussed earlier, this problem can occur when using coordinate-ascent schemes to optimize concave, but not strictly concave, lower bounds.  Other optimization strategies do not suffer from this problem but may have slower rates of convergence.  In this section, we discuss one alternative strategy known as subgradient ascent.  

\begin{algorithm*}[t]
\caption{Subgradient Ascent\label{sga}}
\begin{algorithmic}[1]
\STATE Let $g:\mathbb{R}^n\rightarrow\mathbb{R}$.
\STATE Choose an initial vector $y^0$.

\FOR{$t = 1,2,\ldots$}
\STATE Construct a new vector $y^t$ by setting \[y^t \coloneqq y^{t-1} + \gamma_t h^t\] where $h^t$ is a subgradient of $g$ at $y^{t-1}$ and $\gamma_t$ is the step size at time $t$.
\ENDFOR
\end{algorithmic}
\end{algorithm*}

The subgradient ascent method is a generalization of the gradient ascent method to functions that are not necessarily differentiable.  Let $g:\mathbb{R}^n\rightarrow\mathbb{R}$, and fix a $y^0\in\mathbb{R}^n$.  $h\in\mathbb{R}^n$ is a \textit{subgradient} of $g$ at the point $y^0$ if for all $y\in\mathbb{R}^n$, $g(y) - g(y^0) \geq h\cdot(y - y^0)$.  If $g$ is differentiable, then $\nabla g(y^0)$ is the only subgradient of $g$ at $y^0$.  

The subgradient method to maximize the function $g$ performs the iteration in Algorithm \ref{sga}.  Unlike gradients, subgradients do not necessarily correspond to ascent directions.  However, under certain conditions on the sequence $\gamma_1,\ldots, \gamma_t$, the subgradient algorithm can be shown to converge \cite{shor}. 

The subgradient algorithm can be converted into a distributed algorithm by exploiting the fact that the subgradient of a sum of functions is equal to the sum of the individual subgradients.  Such a strategy has been used to design master/slave algorithms for maximizing the concave lower bounds above \cite{komo07}.  This procedure requires a certain amount of central control, that may not be possible in certain applications.  A double loop method that is equivalent to the subgradient method was proposed in \cite{proximal}.

\section{Graph Covers}
\label{sec:gcover}

Thus far, we have demonstrated that certain parameterized variants of the min-sum algorithm allow us to guarantee both convergence and correctness, upon convergence to locally decodable beliefs.  Even if the beliefs are not locally decodable, they still produce a lower bound on the objective function, but they seem to tell us very little about the argmin of the objective function.  

We have provided relatively little intuition about when we can expect the converged beliefs to be locally decodable.  The success of the min-sum algorithm is intrinsically tied to both the uniqueness of the optimal solution and the ``hardness'' of the optimization problem.  For lower bounds such as those in Section \ref{chp:reparam}, we provide necessary conditions for dual optima to be locally decodable.

These necessary conditions rely on a notion of indistinguishability:  the splitting algorithm, in attempting to solve the
minimization problem on one factor graph, is actually attempting to solve the
minimization problem over an entire family of equivalent (in some sense) factor
graphs. The same notion of indistinguishability has been studied for general distributed message-passing schemes \cite{angluin}, \cite{angluincovers}, and we expect ideas similar to those discussed in this and subsequent sections to be applicable to other iterative algorithms as well.

The above notion of indistinguishability is captured by the formalism of graph covers.  Intuitively, if a graph $H$ covers the graph $G$, then $H$ has the same local structure as $G$. This is potentially problematic as our local message-passing schemes depend only on local graph structure and local potentials.

\begin{definition}
A graph $H$ \textbf{covers} a graph $G$ if there exists a graph homomorphism $h:
H \rightarrow G$ such that $h$ is an isomorphism on neighborhoods (i.e., for all vertices $i\in H$, $\partial i$ is mapped bijectively onto $\partial h(i)$).  If $h(i) = j$, then we say that $i\in H$ is a copy
of $j\in G$.  Further, $H$ is an $\eta$-cover of $G$ if every vertex of $G$ has
exactly $\eta$ copies in $H$.
\end{definition}

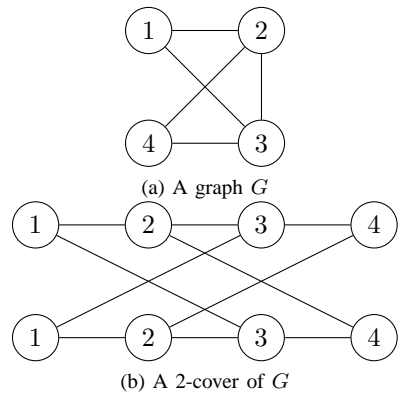
\begin{figure}
\centering
 \subfloat[A graph $G$]{
  \begin{tikzpicture}[scale=1.5]
	\tikzstyle{every node}=[draw,shape=circle];
	\path (0,0) node (X0) {$1$};
	\path (1,0) node (X1) {$2$};
	\path (1,-1) node (X2) {$3$};
	\path (0,-1) node (X3) {$4$};
	
	\draw (X0) -- (X1);
	\draw (X0) -- (X2);
	\draw (X2) -- (X3);
	\draw (X1) -- (X2);
	\draw (X1) -- (X3);
	\end{tikzpicture}
  }                
  \hspace{2cm}
 \subfloat[A 2-cover of $G$]{
  \begin{tikzpicture}[scale=1.5]
	\tikzstyle{every node}=[draw,shape=circle];
	\path (0,0) node (X1) {$1$};
	\path (1,0) node (X2) {$2$};
	\path (2,0) node (X3) {$3$};
	\path (3,0) node (X4) {$4$};
	
	\path (0,-1) node (Y1) {$1$};
	\path (1,-1) node (Y2) {$2$};
	\path (2,-1) node (Y3) {$3$};
	\path (3,-1) node (Y4) {$4$};

	\draw (X1) -- (X2);
	\draw (X2) -- (X3);
	\draw (X3) -- (X4);
	
	\draw (X1) -- (Y3);
	\draw (X2) -- (Y4);
	
	\draw (Y1) -- (Y2);
	\draw (Y2) -- (Y3);
	\draw (Y3) -- (Y4);
	
	\draw (Y1) -- (X3);
	\draw (Y2) -- (X4);
		
	\end{tikzpicture}
  }                 
\caption[An example of a graph cover.]{An example of a graph cover.  Nodes in the cover are labeled by the node that they are a copy of in $G$.}
\label{fig:gcoverex}
\end{figure}

Graph covers may be connected (i.e., there is a path between every pair of vertices) or disconnected.  However, when a graph cover is disconnected, all of the connected components of the cover must themselves be covers of the original graph.  For a simple example of a connected graph cover, see Figure \ref{fig:gcoverex}.  

Every finite cover of a connected graph is an $\eta$-cover for some integer $\eta$.  For every base graph $G$, there exists a graph, possibly infinite, which covers all finite, connected covers of the base graph.  This graph is known as the universal cover.  Throughout this work, we will be primarily concerned with finite, connected covers.

To any finite cover, $H$, of a factor graph $G$ we can associate a collection of potentials derived from the base graph; the
potential at node $i\in H$ is equal to the potential at node $h(i) \in G$.  Together, these potential functions define a new objective function for the factor graph $H$.  In the sequel, we will use superscripts to specify that a particular object is over the factor graph $H$.  For example, we will denote the objective function corresponding to a factor graph $H$ as $f^H$, and we will write $f^G$ for the objective function $f$.

Graph covers, in the context of graphical models, were originally studied in relation to local message-passing algorithms \cite{vontobel}.  Synchronous local message-passing algorithms such as the min-sum and splitting algorithms are incapable of distinguishing the two factor graphs $H$ and $G$ given that the initial messages to and from each node in $H$ are identical
to the nodes they cover in $G$: for every node $i\in G$ the
messages received and sent by this node at time $t$ are exactly the same as the
messages sent and received at time $t$ by any copy of $i$ in $H$.  As a result,
if we use a local message-passing algorithm to deduce an assignment for $i$, then the algorithm
run on the graph $H$ must deduce the same assignment for each copy of $i$.  A similar argument can be made for any sequence of message updates.

Now, consider an objective function $f$ that factors with respect to $G = (V^G, \mathcal{A}^G)$.  For any finite cover $H = (V^H, \mathcal{A}^H)$ of $G$ with covering homomorphism $h:H\rightarrow G$, we can ``lift'' any vector of beliefs, $b^G$, from $G$ to $H$ by defining a new vector of beliefs, $b^H$, such that
\begin{itemize}
\item For all variable nodes $i\in V^H$, $b^H_i = b^G_{h(i)}$.
\item For all factor nodes $\alpha\in \mathcal{A}^H$, $b^H_\alpha = b^G_{h(\alpha)}$.
\end{itemize}
Analogously, we can lift any assignment $x^G$ to an assignment $x^H$ by setting $x^H_i = x^G_{h(i)}$.  

\subsection{Pseudo-codewords}
\label{sec:pseudo}
Surprisingly, minima of the objective
function $f^H$ need not be lifts of the minima of the objective function
$f^G$.  Even worse, the minimum value of $f^G$ does not necessarily correspond to the minimum value of $f^H$. 
This idea is the basis for the theory of pseudo-codewords in the LDPC (low-density parity-check) community
\cite{vontobel}, \cite{vontitt}.  In this community, valid codewords, assignments satisfying a specific set of constraints, on graph covers that are not lifts of
valid codewords of the base graph are referred to as pseudo-codewords.  

\begin{figure}
\centering
 \subfloat[A graph $G$]{
  \begin{tikzpicture}[scale=1]
	\tikzstyle{every node}=[draw,shape=circle];
	\path (0:1) node (X0) {$3$};
	\path (120:1) node (X1) {$2$};
	\path (240:1) node (X2) {$2$};
	
	\draw (X0) -- (X1);
	\draw (X0) -- (X2);
	\draw (X1) -- (X2);
	\end{tikzpicture}
  }                
  \hspace{2cm}
 \subfloat[A 2-cover of $G$]{
  \begin{tikzpicture}[scale=1]
	\tikzstyle{every node}=[draw,shape=circle];
	\path (0:1) node (X0) {$3$};
	\path (60:1) node (X1) {$2$};
	\path (120:1) node (X2) {$2$};
	\path (180:1) node (X3) {$3$};
	\path (240:1) node (X4) {$2$};
	\path (300:1) node (X5) {$2$};
	
	\draw (X0) -- (X5);
	\draw (X0) -- (X1);
	\draw (X1) -- (X2);		
	\draw (X2) -- (X3);		
	\draw (X3) -- (X4);
	\draw (X4) -- (X5);				
	
	\end{tikzpicture}
  }                 
\caption[An example of a maximum weight independent set problem whose graph cover contains a different solution than the original problem.]{An example of a maximum weight independent set problem and a graph cover whose maximum weight independent set is not a copy of an independent set on the original graph.  The nodes are labeled with their corresponding weights.}
\label{fig:mwiscoverex}
\end{figure}
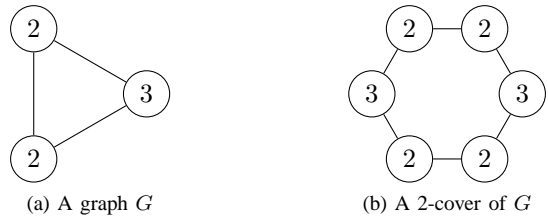

The existence of pseudo-codewords is not unique to coding theory.  Consider the maximum weight independent set problem in Figure \ref{fig:mwiscoverex}.  The maximum weight independent set for the graph in Figure \ref{fig:mwiscoverex} (a) has weight three. The maximum weight independent set on the 2-cover of this graph in Figure \ref{fig:mwiscoverex} (b) has weight seven, which is larger than the lift of the maximum weight independent set from the base graph.  

Because local message-passing algorithms cannot distinguish a factor graph from its covers and our lower bounds provide lower bounds on the objective function corresponding to any finite cover, we expect that maxima of the lower bounds, at best, correspond to the optimal solution on some graph cover of the original problem.  For the splitting algorithm, we can make this intuition precise.
\begin{theorem}
\label{thm:nec}
Let $b$ be a vector of $c$-admissible and min-consistent beliefs for the function
$f$, such that $f$ can be written as a conical combination of the beliefs.  If there exists an assignment $x^G$ that simultaneously minimizes the beliefs, then
\begin{itemize}
\item The assignment $x^G$ minimizes $f^G$, and for any finite cover $H$ of $G$, $x^H$, the lift of $x^G$ to $H$, minimizes $f^H$.

\item For any finite cover $H$ of $G$ with covering homomorphism $h$, if $y^H$ minimizes $f^H$, then for all $i\in H$, $y^H_i \in \arg\min_{x'_i} b_{h(i)}(x'_i)$.
\end{itemize}
\end{theorem}
\begin{IEEEproof}
The beliefs can be lifted from $G$ to $H$.  In other words, the beliefs define a reparameterizaion of the objective function $f^H$.  Consequently, as $f^G$ can be written as a conical combination of the beliefs, $f^H$ can also be written as a conical combination of the beliefs.  The first observation then follows from Corollary \ref{cor:correct}.  For the second observation, observe that, because $x^H$ minimizes $f^H$ and simultaneously minimizes each of the beliefs $b^H$, any other minimum, $y^H$, of $f^H$ must also simultaneously minimize the beliefs $b^H$.  If not, $f(y^H) > f(x^H)$, a contradiction.\\
\end{IEEEproof}

As a consequence of Theorem \ref{thm:nec}, for any choice of the parameter vector that guarantees correctness, the splitting algorithm can only converge to a locally decodable vector of admissible and min-consistent beliefs if the objective function corresponding to any finite graph cover has a unique optimal solution that is a lift of the unique optimal solution on the base graph.

\subsection{Graph Covers and the MAP LP}
\label{sec:gcmap}
The existence of pseudo-codewords is problematic because local message-passing algorithms cannot distinguish a graph from its covers.  For objective functions over finite state spaces, we can relate the previous observations to the MAP LP.  There is a one-to-one correspondence between the rational feasible points of the MAP LP and assignments on graph covers: every rational feasible point of the MAP LP corresponds to an assignment on some cover, and every assignment on a cover of the base graph corresponds to a rational feasible point of the MAP LP.  Similar ideas have been explored in the coding community:  a construction relating psuedo-codewords to graph covers was provided in \cite{vontobel} and an LP relaxation of the codeword polytope, the convex hull of valid codewords, is described in \cite{becpseudo}.

\begin{theorem}
\label{thm:cover}
The vector $\mu$ is a rational, feasible point of the MAP LP for $f^G$ if and only if there exists an $\eta$-cover $H$ of $G$ and an assignment $y^H$ such that
\begin{align*}
f^H(y^H) =& \eta \Big[\sum_{i\in V^G} \sum_{x_i} \mu_i(x_i) \phi_{i}(x_i)\\
&\: + \sum_{\alpha\in \mathcal{A}^G} \sum_{x_\alpha}\mu_\alpha(x_\alpha) \psi_{\alpha}(x_\alpha)\Big].
\end{align*}
\end{theorem}
\begin{IEEEproof}
See Appendix \ref{ap:cover}.\\
\end{IEEEproof}

Because the polyhedron corresponding to the MAP LP is rational, the optimum of the MAP LP is attained at a rational point.  Hence, there is an $\eta$-cover $H$ of $G$ and an assignment $y^H$ that, in the sense of Theorem \ref{thm:cover}, achieves the optimal value of the MAP LP.  This assignment corresponds to a lower bound on the minimum value of any finite cover of $G$ (i.e., for every $\eta'$-cover $H'$ of $G$ and any assignment $y^{H'}$ on $H'$, $\frac{f^H(y^H)}{\eta} \leq \frac{f^{H'}(y^{H'})}{\eta'}$).

The polytope corresponding to the MAP LP is related to the fundamental polytope defined in \cite{vontobel} and the linear program relaxation of \cite{becpseudo}.  The fundamental polytope of \cite{vontobel} contains only the information corresponding to the variable nodes (i.e., each $\mu_i(x_i)$) in the MAP LP whereas the polytope $\mathcal{Q}$ of \cite{becpseudo} is equivalent to the MAP LP for the coding problem.

\subsection{Lower Bounds and the MAP LP}
The relationship between the MAP LP and graph covers allows us to extend the necessary conditions for local decodability of Theorem \ref{thm:nec} to lower bounds whose maximum is equal to the minimum of the MAP LP.  For the analysis in this section, we will concentrate on lower bounds as a function of the messages such that $\sup_m LB(m)$ is equal to the optimum value of the MAP LP.  Lower bounds that are dual to the MAP LP such that strong duality holds always satisfy this property, and many of the lower bounds discussed in this work (e.g., those produced by Algorithm \ref{serial}, TRMP, MPLP, max-sum diffusion, etc.) can be shown to have the required property.  

For simplicity, we consider the lower bound related to Algorithm \ref{serial}.  Let $f^G$ factor over $G=(V,\mathcal{A})$.  Restrict the parameter vector $c$ so that $c_i = 1$ for all $i$, $c_\alpha >0$ for all $\alpha$, and $\sum_{\alpha \in\partial i} c_\alpha \leq 1$ for all $i$.  For any vector of messages, $m$, with corresponding beliefs, $b$, recall the following lower bound on the objective function
\begin{align}
\min_x f^G(x) \geq &\: \sum_{i\in V}  \Big[c_i(1 - \sum_{\alpha\in\partial i}c_\alpha)\min_{x_i}b_i(x_i)\Big]\nonumber\\
&\: + \sum_{\alpha\in\mathcal{A}} \Big[c_\alpha \min_{x_\alpha}b_\alpha(x_\alpha)\Big]\\
& \triangleq LB^G(m).
\end{align}

\begin{theorem}
\label{contight}
If $m^G$ maximizes $LB^G$, then the following are equivalent.\begin{enumerate}
\item The MAP LP has an integral optimum.
\item There is an assignment $x^*$ that simultaneously minimizes each term in the lower bound.
\item If $x^G$ minimizes $f^G$, then for any finite graph cover $H$ of $G$, $x^H$, the lift of $x^G$ to $H$, minimizes $f^H$.
\end{enumerate}
\end{theorem}
\begin{IEEEproof}

$(1\rightarrow 2)$
By Theorem \ref{thm:cover}, there exists a 1-cover of $G$ and an assignment $x^G$ such that $x^G$ minimizes $f^G$.  Because $m^*$ maximizes the lower bound, we have that $\min_x f^G(x)  =  LB^G(m^G)$.  Because of this equality, any assignment that minimizes $f^G$ must also simultaneously minimize each term of the lower bound.

$(2\rightarrow 3)$
Suppose that there exists an assignment, $x^G$, that simultaneously minimizes each component of the lower bound.  Let $H$ be a finite cover of $G$, let $x^H$ be the lift of $x^G$ to $H$, and let $m^H$ be the lift of $m^G$ to $H$.  $x^H$ must simultaneously minimize each term of the lower bound $LB^H$ which implies that $f^H(x^H) = LB^H(m^H)$ and that $x^H$ minimizes $f^H$.

$(3\rightarrow 1)$
Suppose that $f^G$ has a minimum, $x^G$, and for any graph cover $H$ of $G$, $f^H$ has a minimum at $x^H$, the lift of $x^G$ to $H$. By Theorem \ref{thm:cover} every rational feasible point corresponds to an assignment on some finite graph cover $H$ of $G$ and vice versa.  Because every cover is minimized by a lift of $x^G$, the MAP LP must have an integral optimum.\\
\end{IEEEproof}

For lower bounds dual to the MAP LP, the equivalence of $1$ and $2$ also follows from Lemmas 6.2 and 6.4 in \cite{sontag}.  In the case that the optimum of the MAP LP is unique and integral, we have the following corollary.

\begin{corollary}
If $m^G$ maximizes $LB^G$, then the following are equivalent.
\begin{enumerate}
\item The MAP LP has a unique and integral optimum, $x^G$.
\item $f^G$ has a unique minimum $x^G$, and for any finite graph cover $H$ of $G$, $f^H$ is uniquely minimized at $x^H$, the lift of $x^G$ to $H$.
\end{enumerate}
\end{corollary}

Notice that these conditions characterize the existence of an $x^*$ that simultaneously minimizes each of the components of the lower bound, but they do not provide a method for constructing such an $x^*$.  As we saw earlier, local decodability is one condition that ensures that we can construct a solution to the inference problem.  As was the case for the splitting algorithm, Theorem \ref{contight} tells us that dual optimal solutions cannot be locally decodable unless every graph cover of the base factor graph has a unique solution that is a lift of the solution on the base graph or, equivalently, that the optimum of MAP LP is unique and integral. In other words, Theorem \ref{contight} provides necessary, but not sufficient, conditions for dual optima to be locally decodable.  

\subsection{Pairwise Binary Graphical Models}
\label{sec:pwb}
In the special case that the state space is binary (i.e., each $x_i$ can only take one of two values) and the factors depend on at most two variables, we can strengthen the results of the previous sections by providing necessary and sufficient conditions for local decodability.  Previous work on pairwise binary graphical models has focused on the
relationship between the converged beliefs and solutions to the MAP LP \cite{wainkol, weissdbcnt}.  In this work, we focus on the relationships between the base factor graph and its 2-covers.  

In the context of graph covers, the most surprising property of
pairwise binary graphical models is that, for any choice of the vector $c$ and any fixed point of the message updates, we can always construct a 2-cover, $H$, and an assignment on this cover, $x^H$, such that $x^H$ simultaneously minimizes the fixed-point beliefs when lifted to $H$.  In the case that that the objective function can be written as a conical combination of the beliefs, this assignment would be a global minimum of the objective function on the 2-cover (corresponding to a rational optimum of the MAP LP).

\begin{theorem}
\label{2tight}
Let $b^G$ be a vector of $c$-admissible and min-consistent beliefs for the objective
function $f^G$ that factors over $G = (V,\mathcal{A})$.
 If the factorization is pairwise binary, then there exists a 2-cover, $H$, of $G$ and an assignment, $y^*$, on that 2-cover such that
$y^*$ simultaneously minimizes the lifted beliefs.
\end{theorem}
\begin{IEEEproof}
Without loss of generality we can assume that $\mathcal{X}_i = \{0,1\}$ for all $i\in V$.  We will
construct a 2-cover, $H$, of the factor graph $G$ and an assignment $y^*$ such
that $y^*$ minimizes $f_H$.  We will index the copies of variable $i\in G$ in
the factor graph $H$ as $(i,1)$ and $(i,2)$.  First, we will construct the
assignment.  If $\arg\min_{x_i} b^G_i(x_i)$ is unique, then set $y^*_{(i,1)} =
y^*_{(i,2)} = \arg\min_{x_i} b^G_i(x_i)$.  Otherwise, set $y^*_{(i,1)} = 0$ and
$y^*_{(i,2)} = 1$.  Now, we will construct a 2-cover, $H$, such that $y^*$
minimizes each of the beliefs.  We will do this edge by edge.  Consider the edge
$(i,j)\in E$.  There are several possibilities.
\begin{enumerate}
\item $b^G_i$ and $b^G_j$ have unique argmins.  In this case, by Lemma \ref{uniq}, $b^G_{ij}$ is minimized
at $b^G_{ij}(y^*_{(i,1)},y^*_{(j,1)})$.  So, we will add the edges $((i,1), (j,1))$ and
$((i,2), (j,2))$ to $H$.  The corresponding beliefs $b^H_{(i,1)(j,1)}$ and $b^H_{(i,2)(j,2)}$ are minimized at $y^*$.
\item $b^G_i$ has a unique argmin and $b^G_j$ is minimized at both $0$ and $1$ (or
vice versa).  In this case, we have $y^*_{(i,1)} = y^*_{(i,2)}$, $y^*_{(j,1)} = 0$,
and $y^*_{(j,2)} = 1$.  By min-consistency, we can conclude that $b^G_{ij}$ is
minimized at $(y^*_{(i,1)}, 0)$ and $(y^*_{(i,1)}, 1)$.  Therefore, we will add the
edges $((i,1), (j,1))$ and $((i,2), (j,2))$ to $H$.
\item $b^G_i$ and $b^G_j$ are minimized at both $0$ and $1$.  In this case, we have
$y^*_{(i,1)} = 0$, $y^*_{(i,2)} = 1$, $y^*_{(j,1)} = 0$, and $y^*_{(j,2)} = 1$.  By
min-consistency, there is an assignment that minimizes $b^G_{ij}(x_i,x_j)$ with $x_i = 0$ and an assignment that minimizes $b^G_{ij}(x_i,x_j)$ with $x_i = 1$.  This means that
$\arg\min_{x_i,x_j} b^G_{ij}(x_i,x_j)$ contains at least one of the sets $\{(0,0), (1,1)\}$
or $\{(0,1), (1,0)\}$.  In the first case, we will add the edges $((i,1), (j,1))$
and $((i,2), (j,2))$ to $H$, and in the second case, we will add $((i,1), (j,2))$ and
$((i,2), (j,1))$ to $H$.
\end{enumerate}
The constructed $H$ and $y^*$ then satisfy the requirements in the statement of the theorem.\\
\end{IEEEproof}

From the construction in Theorem \ref{2tight}, for the case of pairwise binary graphical models, a solution to the MAP LP must always correspond to an optimal assignment on some 2-cover of the base graph.  A similar phenomenon occurs for cycle codes (see Corollary 3.5 of \cite{dreher}).  

Given the construction in Theorem \ref{2tight}, we can explicitly describe necessary and sufficient conditions for the splitting algorithm to converge to locally decodable beliefs.  

\begin{corollary}
Let $c$ satisfy the conditions of Theorem \ref{conv}.  Algorithm \ref{serial} converges to locally decodable beliefs, $b^G$, if and only if for every cover $H$ of $G$, $f^H$ has a unique minimizing assignment.
\label{cor:localdecode}
\end{corollary}

The proof of Corollary \ref{cor:localdecode} can be found in Appendix \ref{ap:localdecode}. Another consequence of Theorem \ref{2tight} is that, in the pairwise binary case, Algorithm \ref{serial} always converges to a vector of messages that maximizes the lower bound (i.e., the coordinate ascent algorithm does not get stuck).  An alternative proof, based on duality, that the coordinate ascent scheme does not become stuck for these models can be found in \cite{wainkol}. 

\section{Local Versus Global}
\label{sec:lvg}
As a consequence of Theorem \ref{thm:nec}, for any choice of the parameter vector that guarantees correctness, the splitting algorithm can only converge to a locally decodable vector of admissible and min-consistent beliefs if the objective function corresponding to any finite graph cover has a unique optimal solution that is a lift of the unique optimal solution on the base graph.  The corresponding result for pairwise binary factorizations is that this condition is both necessary and sufficient.

These results highlight the inherent weaknesses of conical decompositions and, consequently, the dual approach in general:  conical decompositions guarantee the correctness, on every finite cover of the base graph, of any assignment that simultaneously minimizes each of the beliefs.  For many applications, this requirement on graph covers is very restrictive.  This suggests that the convergent and correct strategy outlined in the previous chapters is not as useful as it seems, and for many applications, the standard min-sum algorithm, which only guarantees a form of local optimality on all graph covers, may still be preferred in practice.  In this section, we briefly discuss specific examples of how these issues arise in practice.

\subsection{Quadratic Minimization}
The quadratic minimization problem provides a simple example of how the guarantee of global optimality can be undesirable in practice.  Let $\Gamma\in \mathbb{R}^{n\times n}$ be a symmetric positive definite matrix and $h\in\mathbb{R}^{n}$.  The quadratic minimization problem is to find the $x\in\mathbb{R}^n$ that minimizes $f(x) = \frac{1}{2}x^T\Gamma x - h^Tx$.  The global optimum must satisfy $\Gamma x = h$, and as a result, minimizing a positive definite quadratic function is equivalent to solving a positive definite linear system.  In this case, the min-sum algorithm is usually called GaBP (Gaussian Belief Propagation).  Consider the following definitions.

\begin{definition}
$\Gamma\in \mathbb{R}^{n\times n}$ is \textbf{scaled diagonally dominant} if $\exists w>0 \in \mathbb{R}^{n}$ such that $|\Gamma_{ii}|w_i > \sum_{j\neq i} |\Gamma_{ij}|w_j$.
\end{definition}

\begin{definition}
$\Gamma\in \mathbb{R}^{n\times n}$ is \textbf{walk summable} if the spectral radius $\varrho(|I - D^{-\frac{1}{2}}\Gamma D^{-\frac{1}{2}}|) < 1$.  Here, $D^{-1/2}$ is the diagonal matrix such that $D^{-1/2}_{ii} = \frac{1}{\sqrt{|\Gamma_{ii}}|}$.
\end{definition}

Here, we use $|A|$ to denote the matrix obtained from $A$ by taking the absolute value of every entry.  

For positive definite $\Gamma$, the sufficiency of walk-summability for the convergence of GaBP was demonstrated in \cite{malioutov} while the sufficiency of scaled diagonal dominance was demonstrated in \cite{quadciamac} and \cite{convexciamac}.  In \cite{allerton09} and \cite{ruozzi3}, we showed the following. 
\begin{theorem}
\label{2cover}
Let $\Gamma$ be a symmetric matrix with positive diagonal. The following are equivalent.
\begin{enumerate}
\item $\Gamma$ is walk summable.
\item $\Gamma$ is scaled diagonally dominant.
\item All covers of $\Gamma$ are positive definite.
\item All 2-covers of $\Gamma$ are positive definite.
\end{enumerate}
\end{theorem}
Here, a cover of the matrix $\Gamma$ is the matrix corresponding to a cover of the pairwise graphical model for the factorization \begin{eqnarray}f(x) = \Big[\sum_i \frac{\Gamma_{ii}}{2}x_i^2 - h_ix_i\Big] + \Big[\sum_{i>j} \Gamma_{ij}x_ix_j\Big].\end{eqnarray}

That is, every matrix that is positive definite but not walk summable is covered by a matrix that is not positive definite.  As any choice of the parameter vector satisfying the conditions of Theorem \ref{correct2} must guarantee the correctness of any collection of locally decodable beliefs on all covers, the splitting algorithm cannot converge to a vector of locally decodable beliefs for such a choice of the parameter vector for any matrix in this class.  As a result, the convergent and correct message-passing algorithms would be a poor choice for solving problems in this particular regime even when the objective function is strictly convex.  A similar observation continues to hold for more general convex functions.  

\subsection{Coding Theory}
Empirical studies of the choice of the parameter vector were discussed in the context of LDPC codes in \cite{yedidiadc}.  Here, the authors discuss an algorithm called the Divide and Concur algorithm that is related to the splitting algorithm.  They demonstrate experimentally that the best decoding performance is not achieved at a choice of the parameter vector that guarantees correctness but at a choice of the parameter vector that only guarantees local optimality.  We refer the interested reader to their paper for more details.

\section{Conclusion}
\label{sec:conc}

In this work, we presented a novel approach for the derivation of convergent and correct message-passing algorithms based on a reparameterization framework motivated by ``splitting'' the nodes of a given factor graph.  Within this framework, we focused on a specific, parameterized family of message-passing algorithms.  We provided conditions on the parameters that guarantee the local or global optimality of locally decodable fixed points, and described a simple coordinate-ascent scheme that guarantees convergence.  

In addition, we showed how to connect assignments on graph covers to rational points of the MAP LP.  This approach allowed us to provide necessary conditions for local decodability and to discuss the limitations of convergent and correct message passing.  These results suggest that, while convergent and correct message-passing algorithms have some advantages over the standard min-sum algorithm in theory, algorithms that only guarantee local optimality still have practical advantages in a variety of settings.

\appendices

\section{Proofs}
\subsection{Proof of Theorem \ref{fpts}}
\label{app:fpts}
\begin{theorem*}
For any vector of fixed-point messages, the corresponding beliefs are admissible
and min-consistent
\end{theorem*}
\begin{IEEEproof}
Let $m$ be a fixed point of the message update equations
\begin{align}
m_{i \rightarrow \alpha}(x_i) & =   \kappa + \phi_i(x_i) + \sum_{\beta\in\partial i \setminus \alpha}  m_{\beta\rightarrow i}(x_i)\\
m_{\alpha\rightarrow i}(x_i) & =  \kappa + \min_{x_{\alpha \setminus i}} \Big[\psi_\alpha(x_\alpha) + \sum_{k\in\alpha \setminus i} m_{k\rightarrow\alpha}(x_k)\Big].
\end{align}
First, we will show that $m$ produces min-consistent beliefs.  Take $\alpha\in\mathcal{A}$ and choose some $i\in\alpha$.  Up to an additive constant we can write
\begin{align}
\min_{x_{\alpha\setminus i}} b_\alpha(x_\alpha) = &  \min_{x_{\alpha\setminus i}}\Big[ \psi_\alpha(x_\alpha) + \sum_{k\in\alpha} m_{k\rightarrow\alpha}(x_k)\Big]\\
 =  &\: m_{i\rightarrow\alpha}(x_i) \nonumber\\
 &\:+ \min_{x_{\alpha\setminus i}}\Big[ \psi_\alpha(x_\alpha) + \sum_{k\in\alpha\setminus i} m_{k\rightarrow\alpha}(x_k)\Big]\\
= &\:  m_{i\rightarrow\alpha}(x_i) + m_{\alpha\rightarrow i}(x_i)\\
=  &\: \phi_i(x_i) + \Big[\sum_{\beta\in\partial i \setminus \alpha}  m_{\beta\rightarrow i}(x_i)\hspace{.1cm}\Big]\nonumber\\
&\: + m_{\alpha\rightarrow i}(x_i)\\
= & \: b_i(x_i).
\end{align}

Next, we can check that the beliefs are admissible.  Again, up to an additive constant,
\begin{align}
f(x) = &\: \sum_i \phi_i(x_i) + \sum_\alpha \psi_\alpha(x_\alpha)\\
= &\: \sum_i \big[\phi_i(x_i) + \sum_{\alpha\in\partial i} m_{\alpha\rightarrow i}(x_i)\Big] \nonumber\\
&\: + \sum_\alpha \Big[\psi_\alpha(x_\alpha) -  \sum_{i\in\alpha} m_{\alpha\rightarrow i}(x_i)\Big]\\
= &\: \sum_i b_i(x_i) + \sum_\alpha \Big[\psi_\alpha(x_\alpha) -  \sum_{i\in\alpha} m_{\alpha\rightarrow i}(x_i)\Big]\\
= &\:  \sum_i b_i(x_i) + \sum_\alpha \Big[\psi_\alpha(x_\alpha) - \sum_{i\in\alpha} b_i(x_i)\nonumber\\
&\: + \sum_{i\in\alpha} b(x_i) - \sum_{i\in\alpha}m_{\alpha\rightarrow i}(x_i)\Big]\\
= &\: \sum_i b_i(x_i) + \sum_\alpha \Big[b_\alpha(x_\alpha) - \sum_{i\in\alpha} b_i(x_i)\Big].
\end{align}

\end{IEEEproof}

\subsection{Proof of Lemma \ref{serialcon}}
\label{app:serialcon}
\begin{lemma*}
Suppose $c_i = 1$ for all $i$, and we perform the update for the edge $(j, \alpha)$ as in Algorithm
\ref{serial}.  If the vector of messages is real-valued after the update, then
$b_\alpha$ is min-consistent with respect to $b_j$.
\end{lemma*}
\begin{IEEEproof}
Let $m$ be the vector of messages before the update and let $m^+$  be the vector
of messages after the update. By the definition of Algorithm \ref{serial}, for each $i\in\alpha\setminus j$, up to an additive constant,
\begin{align}
m^+_{i \rightarrow \alpha}(x_i) & =  \phi_i(x_i) - m_{\alpha\rightarrow i}(x_i) + \sum_{\beta\in\partial i}
c_\beta m_{\beta\rightarrow i}(x_i)\\
& =  \phi_i(x_i) - m^+_{\alpha\rightarrow i}(x_i) + \sum_{\beta\in\partial i} c_\beta m^+_{\beta\rightarrow i}(x_i)\\
& = b_i(x_i) - m_{\alpha\rightarrow i}(x_i).
\end{align}
Similarly,
\begin{align}
m^+_{\alpha\rightarrow j}(x_j) = \kappa + \min_{x_{\alpha \setminus j}}
\Big[\frac{\psi_\alpha(x_\alpha)}{c_\alpha} + \sum_{k\in\alpha \setminus j}
m^+_{k\rightarrow\alpha}(x_k)\Big] \label{eq:betaj}
\end{align}
Let $b^+$ be the vector of beliefs produced derived from $m^+$.  With these observations, up to an additive constant,
\begin{align}
\min_{x_{\alpha\setminus j}} b^+_{\alpha}(x_\alpha) = &   \min_{x_{\alpha \setminus
j}} \Bigg[ \frac{\psi_\alpha(x_\alpha)}{c_\alpha}\nonumber\\
&\: + \sum_{k \in \alpha} \Big[b^+_k(x_k)
 - m^+_{\alpha\rightarrow k}(x_k)\Big] \Bigg]\\
= & \:m^+_{\alpha\rightarrow j}(x_j) + \Big[b^+_j(x_j) - m^+_{\alpha\rightarrow j}(x_j)\Big]\label{eq:p1}\\
= & \:b^+_j(x_j)
\end{align}
where \eqref{eq:p1} follows from the observation in \eqref{eq:betaj}. 
\end{IEEEproof}

\subsection{Proof of Theorem \ref{conv}}
\label{app:conv}
\begin{theorem*}
Suppose $c_i = 1$ for all $i$, $c_\alpha > 0$ for all $\alpha$, and
$\sum_{\alpha\in\partial i}c_\alpha \leq 1$ for all $i$.  If the vector of
messages is real-valued after each iteration of Algorithm \ref{serial}, then for all $t > 0$, $LB(m^t) \geq LB(m^{t-1})$.
\end{theorem*}
\begin{IEEEproof}
The message updates performed for the variable node $j$ in each iteration of Algorithm \ref{serial} cannot
decrease the lower bound.  To see this, let $LB_j(m)$ denote the terms in $LB$ that involve the variable $x_j$.
\begin{align}
LB_j(m) \triangleq &\:  (1-\sum_{\beta\in\partial j} c_\beta)\min_{x_j} b_j(x_j)\nonumber\\
&\: + \sum_{\beta\in\partial j} c_\beta \min_{x_{\beta}}b_\beta(x_\beta)
\end{align}
where $b$ is the vector of beliefs generated by the message vector $m$.

We can upper bound $LB_j$ as 
\begin{align}
LB_j(m) \leq &\:  \min_{x_j} \Bigg[(1-\sum_{\beta\in\partial j} c_\beta)b_j(x_j)\nonumber\\
&\: + \sum_{\beta\in\partial j} c_\beta \min_{x_{\beta\setminus j}}b_\beta(x_\beta)\Bigg]\\
=\:  &\min_{x_j} \Bigg[\phi_j(x_j)  +  \sum_{\beta\in\partial j}c_\beta\min_{x_{\beta
\setminus j}}\Big[\frac{\psi_\beta(x_\beta)}{c_\beta} \nonumber\\ 
&\:+ \sum_{k\in\beta\setminus j}
(b_k(x_k) - m_{\beta\rightarrow k}(x_k))\Big]\Bigg] \label{eq:lbj}
\end{align}
where \eqref{eq:lbj} follows by plugging in the definition of the beliefs and collecting like terms.

The upper bound in \eqref{eq:lbj} does not depend on the choice of the messages from
$\beta$ to $j$ for any $\beta\in\partial j$.  As a result, any choice of these
messages for which the inequality is tight must maximize $LB_j$.  Observe that
the upper bound is tight if there exists an $x_j$ that simultaneously minimizes
$b_j$ and $\min_{x_{\beta\setminus j}} b_\beta$ for each $\beta\in\partial j$. 
By Lemma \ref{serialcon}, this is indeed the case after performing the updates
in Algorithm \ref{serial} for the variable node $j$.   As this is the only part
of the lower bound affected by the update, we have that $LB$ cannot decrease.
Let $m$ be the vector of messages before the update for variable $j$ and $m^+$
the vector after the update (see Lemma \ref{serialcon}).  Define $LB_{-j}$ to be the sum of the terms of the
lower bound that do not involve the variable $x_j$.  By definition and the above,
\begin{align}
LB(m) & =  LB_{-j}(m) + LB_j(m)\\
& \leq  LB_{-j}(m) +  LB_j(m^+)\\
& =  LB_{-j}(m^+) + LB_j(m^+)\label{eq:equals}\\
& =  LB(m^+)
\end{align}
where \eqref{eq:equals} follows from the observation that the update has no effect on messages that are not passed into $j$ or out of any $\alpha$ containing $j$. As $LB(m)$ is bounded from above by $\min_x f(x)$, we can conclude that
the value of the lower bound converges.

Finally, the lower bound has converged if no single variable update can improve
the bound.  By the arguments above, this must mean that there exists an $x_j$
that simultaneously minimizes $b_j$ and $\min_{x_{\alpha\setminus j}} b_\alpha$
for each $\alpha\in\partial j$.  These beliefs may or may not be min-consistent. 
Now, if there exists a unique minimizer $x^*$, then $x_j^*$ must simultaneously
minimize $b_j$ and $\min_{x_{\alpha\setminus j}} b_\alpha$ for each
$\alpha\in\partial j$.  From this we can conclude that $x^*$ simultaneously
minimizes all of the beliefs and therefore, using the argument from Corollary
\ref{cor:correct}, must minimize the objective function.
\end{IEEEproof}

\subsection{Proof of Theorem \ref{thm:cover}}
\label{ap:cover}
\begin{theorem*}
The vector $\mu$ is a rational, feasible point of the MAP LP for $f^G$ if and only if there exists an $\eta$-cover $H$ of $G$ and an assignment $y^H$ such that
\begin{align*}
f^H(y^H) =& \eta \Big[\sum_{i\in V^G} \sum_{x_i} \mu_i(x_i) \phi_{i}(x_i)\\
&\: + \sum_{\alpha\in \mathcal{A}^G} \sum_{x_\alpha}\mu_\alpha(x_\alpha) \psi_{\alpha}(x_\alpha)\Big].
\end{align*}
\end{theorem*}

\begin{IEEEproof}
First, suppose $\mu$ is a rational, feasible point of the MAP LP for $f^G$.  Let $\eta$ be the smallest integer such that $\eta\cdot \mu_i(x_i) \in \mathbb{Z}$ for all $i$ and $x_i$ and $\eta\cdot \mu_\alpha(x_\alpha) \in \mathbb{Z}$ for all $\alpha$ and $x_\alpha$.  Construct an $\eta$-cover $H$ of $G$ by creating $\eta$ copies of each variable and factor node of $G$.  We will think of each copied variable as corresponding to a particular assignment.  For example, consider the $\eta$ copies in $H$ of the variable node $i\in G$.  Exactly $\eta\cdot \mu_i(x_i)$ of these copies will be assigned the value $x_i$.  Similarly, each of the $\eta$ copies in $H$ of the factor node $\alpha\in G$ will correspond to a particular assignment:  $\eta\cdot \mu_\alpha(x_\alpha)$ of these copies will be assigned the value $x_\alpha$.

We can now add edges to $H$ in the following way:  for each copy $\alpha'\in H$ of the factor node $\alpha\in G$, we look at its corresponding assignment and connect $\alpha'$ to a subset of the variables in $H$ that are copies of some $i\in\alpha$, not already connected to a copy of the node $\alpha$ in $H$, whose corresponding assignments are consistent with the assignment for $\alpha'$.  Note that there is not necessarily a unique way to do this, but after this process, every copy in $H$ of $i\in\alpha$ will be connected to a copy of $\alpha$ in $H$.  After repeating this for the remaining factors in $G$, we must have that $H$ is an $\eta$-cover of $G$, and the assignment $y^H$, given by the chosen assignment corresponding to each variable node in $H$, must satisfy 
$f^H(y^H) = \eta \Big[\sum_i \sum_{x_i} \mu_i(x_i) \phi_i(x_i) + \sum_\alpha \sum_{x_\alpha} \mu_\alpha(x_\alpha) \psi_\alpha(x_\alpha)\Big]$.

For the other direction, let $H$ be an $\eta$-cover of $G$ with cover homomorphism $h:H\rightarrow G$, and let $y^H$ be an assignment to the variables in $H$.  Define $\mu_i(x_i) = \frac{1}{\eta}\sum_{j\in H, h(j) = i} \mathbbm{1}_{y^H_j = x_i}$.  Observe that $\eta\cdot \mu_i(x_i)$ is then equal to the number of times in the assignment $y^H$ that some copy in $H$ of the variable $i\in G$ is assigned the value $x_i$.  Similarly, define $\mu_\alpha(x_\alpha) = \frac{1}{\eta} \sum_{\beta\in H, h(\beta) = \alpha} \mathbbm{1}_{y^H_\beta = x_\alpha}$.  With these definitions, $\mu_\alpha(x_\alpha)$ is sum consistent (min-consistency with the min replaced by a sum) with respect to $\mu_i(x_i)$ for each $i\in\alpha$.  This means that $\mu$ is feasible for the MAP LP.  Finally, we have
\begin{align}
 f^H(y^H) = &\: \sum_{i\in H} \phi_{h(i)}(y^H_i) + \sum_{\alpha\in H} \psi_{h(\alpha)}(y^H_\alpha)\\
 = &\: \sum_{i\in G} \sum_{x_i} \eta\cdot\mu_i(x_i) \phi_{i}(x_i)\nonumber\\
 &\: + \sum_{\alpha\in G} \sum_{x_\alpha} \eta\cdot \mu_\alpha(x_\alpha)
  \psi_{\alpha}(x_\alpha).
\end{align}
\end{IEEEproof}
\subsection{Proof of Corollary \ref{cor:localdecode}}
\label{ap:localdecode}
\begin{corollary*}
Let $c$ satisfy the conditions of Theorem \ref{conv}.  Algorithm \ref{serial} converges to locally decodable beliefs, $b^G$, if and only if for every cover $H$ of $G$, $f^H$ has a unique minimizing assignment.
\end{corollary*}
\begin{IEEEproof}
The lower bound has converged if no single variable update can improve
the bound.  By the arguments in the proof of Theorem \ref{conv}, this must mean that for each $j$ there exists an $x^*_j$
that simultaneously minimizes $(1-\sum_{i\in\partial j} c_{ij})b_j(x_j)$ and $\min_{x_i} b_{ij}(x_i,x_j)$
for each $i\in\partial j$.  Notice that these beliefs may or may not be min-consistent.  However, as observed in the proof of Theorem \ref{conv}, when $LB_j$ cannot be improved it is independent of the messages passed from $i$ to $j$ for each $i\in\partial j$.  As a result, we may assume that the beliefs are min-consistent as they must have the same minima as the min-consistent beliefs. 

For one direction of the proof, suppose that there exists an $x^*$ such that the beliefs, on $G$, are locally decodable to $x^*$. This implies that for any cover $H'$ of $G$ the lift of $b$ to $H'$ is locally decodable to the lift of $x^*$ to $H'$.  Recall that this must imply that every graph cover has a unique minimizing assignment; the lower bound is tight to each graph cover and this implies that any minimizing assignment must simultaneously minimize the beliefs.

For the other direction, suppose that every graph cover has a unique minimizing assignment.  We can construct a vector $y^*$ and a graph cover $H$ as in Theorem \ref{2tight}:   if there is a unique $x_j$ that simultaneously minimizes $(1-\sum_{i\in\partial j} c_{ij})b_j(x_j)$ and $\min_{x_i} b_{ij}(x_i, x_j)$
for each $i\in\partial j$, then set $y^*_{j_1}$ and $y^*_{j_2}$ equal to this $x_j$.  Otherwise, set $y^*_{j_1} = 0$ and
$y^*_{j_2} = 1$.  The 2-cover, $H$, can then be constructed using the vector $y^*$ as in the proof of the theorem.  Construct a vector $z^*$ similarly to the vector $y^*$ but swapping the role of zero and one:   if there is a unique $x_j$ that simultaneously minimizes $(1-\sum_{i\in\partial j} c_{ij})b_j(x_j)$ and $\min_{x_i} b_{ij}(x_i, x_j)$
for each $i\in\partial j$, then set $z^*_{j_1}$ and $z^*_{j_2}$ equal to this $x_j$.  Otherwise, set $z^*_{j_1} = 1$ and $z^*_{j_2} = 0$.  By the symmetry of the construction, the vector $z^*$ also simultaneously minimizes the beliefs on $H$.

Finally,  as each graph cover has a unique minimizing assignment, we must have $z^* =y^*$ and the result follows.  We have $z^* = y^*$ if and only if the beliefs, on $H$, are locally decodable to $y^*$.   In addition, if $z^* = y^*$, then there exists an $x^*$ such that the beliefs, on $G$, are locally decodable to $x^*$.
\end{IEEEproof}

\bibliographystyle{IEEEtranS}
\bibliography{biblio}

\end{document}